\documentclass[final]{siamltex}
\usepackage[]{amsmath,epsfig}

\newcommand{\mb}[1]{\mbox{\boldmath$#1$}}
\newcommand{\p}{\partial}
\newcommand{\ds}{\displaystyle}
\newcommand{\beq}{\begin{eqnarray}}
\newcommand{\beqq}{\begin{eqnarray*}}
\newcommand{\eeq}{\end{eqnarray}}
\newcommand{\eeqq}{\end{eqnarray*}}
\newcommand{\A}{\mbox{\boldmath$a$}}
\newcommand{\B}{\mbox{\boldmath$B$}}
\newcommand{\x}{\mbox{\boldmath$x$}}
\newcommand{\n}{\mbox{\boldmath$n$}}
\newcommand{\y}{\mbox{\boldmath$y$}}
\newcommand{\J}{\mbox{\boldmath$J$}}

\newcommand{\w}{\mbox{\boldmath$w$}}
\newcommand{\U}{\mbox{\boldmath$v$}}
\font\bb=msbm10 at 12pt

\def\rR{\hbox{\bb R}}

\def\ds#1{\displaystyle{#1}}

\title{Partially Reflected Diffusion}
\author{A. Singer\thanks{Department of Mathematics, Program in Applied Mathematics, Yale University,
10 Hillhouse Ave. PO Box 208283, New Haven, CT 06520-8283
(amit.singer@yale.edu)} \and Z. Schuss\thanks{Department of Applied
Mathematics, Tel-Aviv University, Ramat-Aviv 69978 Tel-Aviv, Israel
(schuss@post.tau.ac.il). The research of this author was partially
supported by a grant from TAU.} \and A. Osipov\thanks{Institute of
Mathematics, The Hebrew University, Jerusalem 91904, Israel
(andreiosipov2@gmail.com)} \and D. Holcman\thanks {Department of
Mathematics, Weizmann Institute of Science, Rehovot 76100 Israel
(david.holcman@weizmann.ac.il). This author is incumbent to the
Madeleine Haas Russell Career Development Chair.}}
\begin{document}

\maketitle

\begin{abstract}
The radiation (reaction, Robin) boundary condition for the diffusion
equation is widely used in chemical and biological applications to
express reactive boundaries. The underlying trajectories of the
diffusing particles are believed to be partially absorbed and
partially reflected at the reactive boundary, however, the relation
between the reaction constant in the Robin boundary condition and
the reflection probability is not well defined. In this paper we
define the partially reflected process as a limit of the Markovian
jump process generated by the Euler scheme for the underlying It\^o
dynamics with partial boundary reflection. Trajectories that cross
the boundary are terminated with probability $P\sqrt{\Delta t}$ and
otherwise are reflected in a normal or oblique direction. We use
boundary layer analysis of the corresponding master equation to
resolve the non-uniform convergence of the probability density
function of the numerical scheme to the solution of the
Fokker-Planck equation in a half space, with the Robin constant
$\kappa$. The boundary layer equation is of the Wiener-Hopf type. We
show that the Robin boundary condition is recovered if and only if
trajectories are reflected in the co-normal direction
$\mb{\sigma}\mb{n}$, where $\mb{\sigma}$ is the (possibly
anisotropic) constant diffusion matrix and $\n$ is the unit normal
to the boundary. Otherwise, the density satisfies an oblique
derivative boundary condition. The constant $\kappa$ is related to
$P$ by $\kappa= rP\sqrt{\sigma_n}$, where $r=1/\sqrt{\pi}$ and
$\sigma_n=\n^T\mb{\sigma}\n$. The reflection law and the relation
are new for diffusion in higher-dimensions. %The constant $r$ for the
%Euler scheme is not the same as that for other schemes, e.g., for a
%discrete random walk with radiation boundaries, $r=1/\sqrt{2}$.
\end{abstract}



\section{\label{sec:intro}Introduction}
The Fokker-Planck equation (FPE) with radiation (also called
reactive or Robin) boundary conditions is widely used to describe
diffusion in a biological cell with chemical reactions on its
surface \cite{LammSchultenII}, \cite{Zwanzig}, \cite{McCammon1},
\cite{BBS}, \cite{BMZS}, \cite{Monine}, \cite{McCammonRev},
\cite{ErbanChapman}, \cite{AndrewsBray}. The Robin boundary
conditions are used in \cite{Zwanzig}, \cite{BBS}, \cite{BMZS},
\cite{Monine} as a homogenization of mixed Dirichlet-Neumann
boundary conditions given on scattered small absorbing windows in an
otherwise reflecting boundary. The latter may represent, e.g.,
ligand binding or pumping out ions at sites on the boundary of a
biological cell and no flux through the remaining boundary. The
reactive rate constant in the Robin boundary conditions is chosen in
the homogenization process so that the decay rate of the survival
probability is the same as that in the mixed Dirichlet-Neumann
boundary value problem.

The definition of the It\^o stochastic dynamics
 \beq
 \dot
x=a(x,t)+\sqrt{2\sigma(x,t)}\,\dot w,\label{SDE}
 \eeq
on the positive axis with total or partial reflection at the origin
was given first by Feller \cite{Feller} for the one-dimensional case
with $a(x,t)$ and $\sigma(x,t)$ independent of $t$, as a limit of
It\^o processes, which are terminated when they reach the boundary
or moved instantaneously to a point $x=\rho_j>0$ with probability
$p_j$. When $p_j\to1$ and $\rho_j\to0$ with
 \beq
\lim_{j\to\infty}\frac{1-p_j}{\rho_j}=c,\label{c}
 \eeq
where $c$ is a constant, the partially reflected process converges
to a limit. The transition probability density function (pdf) of the
limit process, $\mbox{$p(y,t\,|\,x,s)\,dy=\Pr\{x(t)\in
(y,y+dy)\,|\,x(s)=x\}$}$, is the solution of the FPE
 \beq
\frac{\p p(y,t\,|\,x,s)}{\p t} =-\frac{\p [a(y,t)p(y,t\,|\,x,s)]}{\p
y}+\frac{\p^2 [\sigma(y,t)p(y,t\,|\,x,s)]}{\p y^2},\label{FPE}
 \eeq
or equivalently,
 \beqq
\frac{\p p(y,t\,|\,x,s)}{\p t}=-\frac{\p J(y,t\,|\,x,s)}{\p y}
\quad\mbox{for all} \quad y,x>0,
 \eeqq
where
 \beq
J(y,t\,|\,x,s)=a(y,t)p(y,t\,|\,x,s)-\frac{\p[\sigma(y,t)p(y,t\,|\,x,s)]}{\p
y},\label{J}
 \eeq
is the flux. The initial condition is
 \beq
p(y,t\,|\,x,s)\to \delta(y-x)\quad\mbox{as}\quad t\downarrow
s\label{IC}
 \eeq
and the radiation boundary condition is
 \beq
-J(0,t\,|\,x,s)=\kappa p(0,t\,|\,x,s),\label{radiation}
 \eeq
where $\kappa$ is a constant related to the constant $c$ and to the
values of the coefficients at the boundary. The no flux and
Dirichlet boundary conditions are recovered if $c=0$ or $c=\infty$,
respectively. Feller's method does not translate into a Brownian
dynamics simulation of the limit process, because his approximations
are continuous-time It\^o processes. Skorokhod \cite{Skorokhod1}
defines the reflection process inside the boundary. Several
numerical schemes have been proposed for simulating this process
(see, e.g., \cite{Skorokhod1}, \cite{Asmussen}, \cite{Lepingle},
\cite{Constantini}). The main issue there is to approximate the
local time spent on the boundary.

The definition of a diffusion process with absorbing or reflecting
boundaries as limits of Markovian jump processes, which is the basis
for all simulations, gives in the limit diffusion processes with
well defined boundary behavior. However, the definition of a
diffusion process with partially reflecting boundaries as a limit of
Markovian jump processes gives different diffusions for different
jump processes. This is expressed in different relations between the
termination probability of the jump process and the boundary
conditions for the Fokker-Planck equations (see, e.g.,
\cite{ErbanChapman}). The process $x(t)$ defined by equation
(\ref{SDE}) with partially absorbing boundaries can be defined as
the limit of the solutions of the Markovian jump processes generated
by the Euler scheme
  \beq
 x_{\Delta t}(t+\Delta t)&=&x_{\Delta t}(t)+a(x_{\Delta t}(t),t)\Delta
t+\sqrt{2\sigma(x_{\Delta t}(t),t)}\,\Delta
 w(t,\Delta t)\quad\mbox{for}\quad t\geq s\label{Euler}\\
 &&\nonumber\\
x_{\Delta t}(s)&=&x\label{ICE}
 \eeq
in the interval $x>0$, for  $0\leq t-s\leq T$, with $\Delta t=T/N,\
t-s = iT/N\ (i=0,1,\dots,N)$, where for each $t$ the random
variables $\Delta w(t,\Delta t)$ are normally distributed and
independent with zero mean and variance $\Delta t$. The partially
absorbing boundary condition for (\ref{Euler}) has to be chosen so
that the pdf $p_{\Delta t}(x,t)$ of $x_{\Delta t}(t)$ converges to
the solution of (\ref{FPE})-(\ref{radiation}).  At a partially
reflecting boundary for (\ref{Euler}), the trajectories are
reflected with probability (w.p.) $R$ and otherwise terminated
(absorbed), once they cross the origin. We show below that keeping
$R$ constant (e.g., $R=1/2$) as $\Delta t\to 0$ leads to the
convergence of the pdf $p_{\Delta t}(x,t)$ to the solution of the
FPE with an absorbing rather than the Robin boundary condition. Thus
the reflection probability $R$ must increase to $1$ as $\Delta t \to
0$ in order to yield the Robin condition (\ref{radiation}).
Moreover, the reactive constant $\kappa$ is related to the limit
\begin{equation}
\lim_{\Delta t\to 0} \frac{1-R}{\sqrt{\Delta t}}=P.
\end{equation}
The reflecting boundary condition is recovered for $P=0$, while the
absorbing boundary condition is obtained for $P=\infty$. Motivated
by these considerations, we design the following simple boundary
behavior for the simulated trajectories that cross the boundary,
identified by $x_{\Delta t}(t)+a(x_{\Delta t}(t),t)\Delta
t+\sqrt{2\sigma(x_{\Delta t}(t),t)}\,\Delta
 w<0$,
 \beq
 x_{\Delta t}(t+\Delta t)=\left\{
\begin{array}{l}
-(x_{\Delta t}(t)+a(x_{\Delta t}(t),t)\Delta
t+\sqrt{2\sigma(x_{\Delta t}(t),t)}\,\Delta
 w)\quad\mbox{w.p.}\ 1-P\sqrt{\Delta
t}\\
 \label{design}\\
 \mbox{terminate trajectory otherwise}.
\end{array}\right.
  \eeq
The exiting trajectory is normally reflected w.p.
 \beq
 R=1-P\sqrt{\Delta t}\label{RDP}
 \eeq
and is otherwise terminated (absorbed). The scaling of the
termination probability with $\sqrt{\Delta t}$ reflects the fact
that the discrete unidirectional diffusion current at any point,
including the boundary, is $O\left(1/\sqrt{\Delta t}\right)$ (see
\cite{unidirect}, \cite{unidir}). This means that the number of
discrete trajectories hitting or crossing the boundary in any finite
time interval increases as $1/\sqrt{\Delta t}$. Therefore, to keep
the efflux of trajectories finite as $\Delta  t\to0$, the
termination probability of a crossing trajectory, $1-R$, has to be
$O(\sqrt{\Delta t})$. The pdf $p_{\Delta t}(x,t)$, however, does not
converge to the solution $p(x,t)$ of (\ref{FPE})-(\ref{radiation})
on the boundary, as discussed in Section \ref{sec:anal}. This is due
to the formation of a boundary layer, as is typical for diffusion
approximations of Markovian jump processes that jump over the
boundary \cite{KMST1}, \cite{KMST2}, \cite{KMST3}. The boundary
layer equations are typically Wiener-Hopf integral equations. The
Wiener-Hopf boundary layer equation for the particular case of a
partially reflected Brownian motion on the positive axis (i.e.,
$a(x,t)=0$ and $\sigma(x,t)=\sigma$ in (\ref{Euler})) was recently
solved in \cite{ErbanChapman} and the relationship
$\kappa=P\sqrt{\sigma}/\sqrt{\pi}$ was found.

The convergence of the pdf of an Euler scheme has been studied in
\cite{Gobet}, \cite{Talay} for the higher-dimensional problem with
oblique reflection. Bounds on the integral norm of the approximation
error are given for the solution of the backward Kolmogorov
equation. These, however, do not resolve the boundary layer of the
pdf of the numerical solution. The solution of the forward equation
for the Euler scheme converges non-uniformly to the solution of the
Fokker-Planck equation due to the appearance of a boundary layer in
the first order spatial derivative. This distorts the boundary flux
and gives incorrect boundary conditions. A boundary layer expansion
is needed to capture the boundary phenomena.

The derivation of the radiation condition has a long history.
Collins and Kimball \cite{CollinsKimball} (see also \cite{Goodrich})
derived the radiation boundary condition (\ref{radiation}) for the
limit $p(x,t)=\lim_{\Delta t \to 0}p_{\Delta t}(x,t)$ from an
underlying discrete random walk model on a semi-infinite
one-dimensional lattice with partial absorbtion at the endpoint.
Their model assumes constant diffusion coefficient and vanishing
drift, for which they find the reactive constant in terms of the
absorbtion probability and the diffusion coefficient. Previous
simulation schemes that recover the Robin boundary condition
\cite{LammSchultenII}, \cite{Ghoniem}, \cite{Northrup1986},
\cite{Ladd2004}, \cite{Green1988}  make use of the explicit solution
to the half space FPE with linear drift term and constant diffusion
coefficient with a Robin condition. In \cite[and references
therein]{Ladd} the specular reflection method near a reflecting
boundary has been shown to be superior to other methods, such as
rejection, multiple rejection and interruption.

An apparent paradox arises when using (\ref{Euler}) and other
schemes: while the pdf $p_{\Delta t}(y,t\,|\,x,s)$ of the solution
of (\ref{Euler}), (\ref{ICE}), (\ref{design}), (\ref{RDP}) converges
to the solution of the FPE (\ref{FPE}) and the initial condition
(\ref{IC}), each approximant $p_{\Delta t}(y,t\,|\,x,s)$ does not
satisfy the boundary condition (\ref{radiation}), not even
approximately, that is, the error does not decay as $\Delta t\to0$.
For a general diffusion coefficient and drift term, the boundary
condition is not satisfied even for the case of a reflecting
boundary condition. This problem plagues other schemes as well. The
apparent paradox is due to the non-uniform convergence of $p_{\Delta
t}(y,t\,|\,x,s)$ to the solution $p(y,t\,|\,x,s)$ of the
Fokker-Planck equation, caused by a boundary layer in $p_{\Delta
t}(y,t\,|\,x,s)$, as is typical of boundary behavior of diffusion
approximations to Markovian jump processes. The limit
$p(y,t\,|\,x,s)$, however, satisfies the boundary condition
(\ref{radiation}) for some $\kappa$. Our analysis can be extended to
other schemes in a straightforward way. It is well known that the
Euler scheme produces an $O(\sqrt{\Delta t})$ error in estimating
the mean first passage time to reach an absorbing boundary. There
are several recipes to reduce the discretization error to $O(\Delta
t)$ \cite{Beccaria}, \cite{Honerkamp}, \cite{Mannella2},
\cite{Mannella1}, \cite{CliffordGreen}. Another manifestation of the
boundary layer is that the approximation error of the pdf near
absorbing or reflecting boundaries is $O(\sqrt{\Delta t})$, and
methods, including \cite{LammSchultenII}, \cite{Peters}  reduce this
error to $O(\Delta t)$. Thus, we expect the formation of a boundary
layer of size $O(\sqrt{\Delta t})$ for the Euler scheme
(\ref{Euler}) with the boundary behavior (\ref{design}).

This paper is concerned with the convergence of the partially
reflecting Markovian jump process generated by (\ref{Euler}),
(\ref{design}) in one and higher dimensions. We show that this
scheme, with the additional requirement that the pdf converges to
the solution of the FPE with a given Robin boundary condition,
defines a unique diffusion process with partial reflection at the
boundary. This definition is then generalized to higher dimensions.
In contrast to the Collins and Kimball \cite{CollinsKimball}
discrete scheme, this definition is not restricted to lattice points
and the drift and diffusion coefficients may vary. The advantage of
the current suggested design (\ref{design}) is its simplicity, which
is both easily and efficiently implemented and amenable to analysis.
There is no need to make any assumptions on the structure of the
diffusion coefficient or the drift. From the theoretical point of
view, it serves as a physical interpretation for the behavior of
diffusive trajectories near a reactive boundary.

Our main result in the one-dimensional case is the relation between
the reactive "constant" $\kappa(t)$ and the absorbtion parameter $P$
for the dynamics (\ref{SDE}) on the positive axis with drift and
with a variable diffusion coefficient,
 \beq
 \kappa(t)= rP\sqrt{\sigma(0,t)},\quad r=\frac{1}{\sqrt{\pi}}.\label{kP}
 \eeq
The relation (\ref{kP}) is new for diffusion with variable
coefficients. The value $r=1/\sqrt{\pi}$ is different than values
obtained for other schemes, e.g., than the value $r=1/\sqrt{2}$,
predicted by the discrete random walk theory of radiation boundaries
\cite{CollinsKimball}. Values of $r$ for other schemes are given in
\cite{ErbanChapman}. We show the effect of using (\ref{kP}) in
numerical simulations.

The scheme (\ref{design}) is generalized to diffusion with drift and
anisotropic constant diffusion matrix $\mb{\sigma}(t)$ in the half
space, $x_1>0$, with partial oblique reflection. We show that the
Robin boundary condition is recovered if and only if trajectories
are reflected in the direction of the unit vector
 \beq
 \U=\frac{\mb{\sigma}\mb{n}}{\|\mb{\sigma}\mb{n}\|},
 \eeq
where $\n$ is the unit normal to the boundary. The radiation
parameter $\kappa(\x,t)$ in the $d$-dimensional Robin boundary
condition and the absorbtion parameter $P(\x)$ are related by
 \beq
 \kappa(\x,t)= rP(\x)\sqrt{\sigma_n(t)},\quad x_1=0,\label{kP1}
 \eeq
with $r$ given in (\ref{kP}) and $\sigma_n(t)=\n^T\mb{\sigma}(t)\n$.
The relation (\ref{kP1}) is new for higher-dimensional diffusion in
a half space with drift and anisotropic diffusion matrix.

In the most common case of constant isotropic diffusion our result
extends to domains with curved boundaries. This is due to the fact
that a smooth local mapping of the domain to a half space with an
orthogonal system of coordinates preserves the constant isotropic
diffusion matrix, though the drift changes according to It\^o's
formula. In this case the vector $\U$ coincides with the normal
$\n$.

%A similar boundary layer analysis can be applied to many numerical
%schemes with different convergence rates, e.g.,
%\cite{LammSchultenII}, that approximate the partially reflected
%diffusion process. The more general problem of variable diffusion
%matrix in a domain with a curved boundary will be discussed in a
%separate paper.

\section{\label{sec:anal}Boundary layer analysis in one dimension}
The aim of the boundary layer analysis below is to examine the
convergence of the pdf $p_{\Delta t}(y,t\,|\,x,s)$ of the solution
$x_{\Delta t}(t)$ of (\ref{Euler}), (\ref{ICE}) to the solution
$p(y,t\,|\,x,s)$ of (\ref{FPE})-(\ref{radiation}), and to find the
relation between the parameter $P$ of (\ref{design}) and the
reactive constant $\kappa$ in (\ref{radiation}). Using abbreviated
notation, the pdf $p_{\Delta t}(y,t\,|\,x,s)=p_{\Delta t}(y,t)$
satisfies the forward Kolmogorov equation \cite{unidirect},
\cite{unidir}, \cite{KMST1}, \cite{KMST2},\cite{KMST3},
\cite{Keller}
 \beq
 p_{\Delta t}(y,t+\Delta
t)&=&\int_0^\infty\frac{p_{\Delta t}(x,t)}{\sqrt{4\pi
\sigma(x,t)\Delta t}}
\left\{\exp\left[-\ds{\frac{\left(y-x-a(x,t)\Delta
 t\right)^2}{4\sigma(x,t)\Delta t}}\right]+\right.\nonumber\\
&&\nonumber\\
 &&\left.(1-P\sqrt{\Delta
t})\exp\left[-\ds{\frac{\left(y+x+a(x,t)\Delta
 t\right)^2}{4\sigma(x,t)\Delta
t}}\right]\right\}\,dx\label{prop}.
 \eeq
For $P=0$ the pdf $p_{\Delta t}(y,t)$ satisfies the boundary
condition
 \beq
 \frac{\p p_{\Delta t}(0,t)}{\p y}=0,\label{px0}
 \eeq
which is obtained by differentiation of (\ref{prop}) with respect to
$y$ at $y=0$. If $P\neq0$, we obtain
 \beq
 \frac{\p p_{\Delta t}(0,t+\Delta t)}{\p
y}=\frac{p_{\Delta t}(0,t)P}{\sqrt{4\pi\sigma(0,t)}}+O(\sqrt{\Delta
t}),
 \label{px11}
 \eeq
which holds also in the limit $\Delta t\to 0$.
 %\beq
% \frac{\p p(0,t)}{\p
% y}=\frac{p(0,t)P}{\sqrt{4\pi\sigma(0,t)}}.
% \label{px1}
% \eeq
However, the order of the limits $\Delta t\to0$ and $y\downarrow0$
matters, indeed,
 \beq
 \lim_{\Delta t\to0}\lim_{y\downarrow0}\frac{\p p_{\Delta t}(y,t)}{\p
y}\neq\lim_{y\downarrow0}\lim_{\Delta t\to0}\frac{\p p_{\Delta
t}(y,t)}{\p y}.
 \eeq
The limit of (\ref{px11}) is not the boundary condition that the
limit function $p(y,t)=\ds\lim_{\Delta t\to 0}p_{\Delta t}(y,t)$
(for $y>0$) satisfies. To find the boundary condition of $p(y,t)$,
in either case $P=0$ or $P\neq0$, we show below that $p(y,t)$
satisfies the FPE (\ref{FPE}) and the initial condition (\ref{IC})
for all $y>0$. Since for $P=0$ the simulation preserves probability
(the population of trajectories),
 \beq
0=\frac{d}{dt} \int_0^\infty p(x,t)\,dx=-\frac{\p
[\sigma(0,t)p(0,t)]}{\p y} +a(0,t)p(0,t)=J(0,t).\label{flux}
 \eeq
Equation (\ref{flux}) is the no-flux boundary condition. The
discrepancy between (\ref{flux}) and (\ref{px0}) is due to the
nonuniform convergence of $p_{\Delta t}(y,t)$ to its limit $p(y,t)$
in the interval. There is a boundary layer of width $O(\sqrt{\Delta
t})$, in which the boundary condition (\ref{px0}) for $p_{\Delta
t}(y,t)$ changes into the boundary condition (\ref{flux}) that
$p(y,t)$ satisfies. To analyze the discrepancy between (\ref{px0})
and (\ref{flux}), we introduce the local variable $y=\eta
\sqrt{\Delta t}$ and the boundary layer solution
\begin{equation}
p_{BL}(\eta,t) = p_{\Delta t}(\eta\sqrt{\Delta t},t).
\end{equation}
Changing variables $x=\xi\sqrt{\Delta t}$ in the integral
(\ref{prop}) gives
\begin{eqnarray}
 p_{BL}(\eta,t+\Delta
t)&=&\int_0^\infty\frac{p_{BL}(\xi,t)}{\sqrt{4\pi
\sigma(\xi\sqrt{\Delta t},t)}}
\left\{\exp\left[-\ds{\frac{\left(\eta-\xi-a(\xi\sqrt{\Delta
t},t)\sqrt{\Delta t}\right)^2}{4\sigma(\xi\sqrt{\Delta t},t)}}\right]+\right.\nonumber\\
&&\nonumber\\
 &&\left.(1-P\sqrt{\Delta
t})\exp\left[-\ds{\frac{\left(\eta+\xi+a(\xi\sqrt{\Delta
t},t)\sqrt{\Delta
 t}\right)^2}{4\sigma(\xi\sqrt{\Delta t},t)}}\right]\right\}\,d\xi\label{BLint}.
\end{eqnarray}
The boundary layer solution has an asymptotic expansion in powers of
$\sqrt{\Delta t}$
\begin{equation}
p_{BL}(\eta,t) \sim p_{BL}^{(0)}(\eta,t) + \sqrt{\Delta
t}\,p_{BL}^{(1)}(\eta,t) + \Delta t \, p_{BL}^{(2)}(\eta,t) +
\ldots.\label{pBL}
\end{equation}
Expanding all functions in (\ref{BLint}) in powers of $\sqrt{\Delta
t}$ and equating similar orders, we obtain integral equations that
the asymptotic terms of (\ref{pBL}) must satisfy. The leading order
$O(1)$ term gives the Wiener-Hopf-type equation on the half line
\begin{equation}
p_{BL}^{(0)}(\eta,t) = \int_0^{\infty}
\frac{p_{BL}^{(0)}(\xi,t)}{\sqrt{4\pi\sigma(0,t)}}\left\{\exp\left[-\frac{(\eta-\xi)^2}{4\sigma(0,t)}
\right] + \exp\left[-\frac{(\eta+\xi)^2}{4\sigma(0,t)} \right]
\right\}\,d\xi,\label{pBL0-int}
\end{equation}
for $\eta>0$. The kernel
\begin{equation}
K(\eta,\xi) = \exp\left[-\frac{(\eta-\xi)^2}{4\sigma(0,t)} \right] +
\exp\left[-\frac{(\eta+\xi)^2}{4\sigma(0,t)}\right]
\end{equation}
is an even function of $\eta$ and $\xi$, i.e. $K(\eta,\xi) =
K(-\eta,\xi) = K(\eta,-\xi) = K(-\eta,-\xi)$. Therefore, we extend
$p_{BL}^{(0)}(\xi,t)$ to the entire line as an even function
($p_{BL}^{(0)}(\xi,t)=p_{BL}^{(0)}(-\xi,t)$), and rewrite
(\ref{pBL0-int}) as
\begin{equation}
p_{BL}^{(0)}(\eta,t) = \int_{-\infty}^{\infty}
\frac{p_{BL}^{(0)}(\xi,t)}{\sqrt{4\pi\sigma(0,t)}}\exp\left[-\frac{(\eta-\xi)^2}{4\sigma(0,t)}
\right]\,d\xi,\label{pBL0-int2}
\end{equation}
for $-\infty < \eta < \infty$. The only solution of the integral
equation (\ref{pBL0-int2}) is the constant function, that is,
$p_{BL}^{(0)}(\eta,t)=f(t)$, independent of $\eta$. This follows
immediately from the Fourier transform of (\ref{pBL0-int2}), whose
right hand side is a convolution.

Away from the boundary layer the solution admits an outer solution
expansion
\begin{equation}
p_{OUT}(y,t) \sim p_{OUT}^{(0)}(y,t) + \sqrt{\Delta
t}p_{OUT}^{(1)}(y,t) + \ldots,
\end{equation}
where $p_{OUT}^{(0)}$ satisfies the Fokker-Planck equation
(\ref{FPE}) and the initial condition (\ref{IC}). Indeed, the
integrals in (\ref{prop}) are of Laplace type with the small
parameter $\Delta t$. For interior points $y\gg\sqrt{\Delta t}$ the
second integral, which represents only boundary interactions, is
negligible relative to the first one. We change variables in
(\ref{prop}) by setting
 \beqq
\eta=\frac{y-x-a(x,t)\Delta
 t}{\sqrt{2\sigma(x,t)\Delta t}},
 \eeqq
and extend integration over the entire line in the first integral
and expand all functions in powers of $\sqrt{\Delta t}$. The
resulting integrals are moments of the normal distribution. We
obtain
 \beqq
 \frac{p_{\Delta t}(y,t+\Delta t)-p_{\Delta t}(y,t)}{\Delta t}=-\frac{\p [a(y,t)p_{\Delta t}(y,t)]}{\p
y}+\frac{\p^2 [\sigma(y,t)p_{\Delta t}(y,t)]}{\p y^2}+O(\sqrt{\Delta
t}).
 \eeqq
The leading term in the expansion of $p_{\Delta t}(y,t)$ is
$p_{OUT}^{(0)}(y,t)$, which therefore satisfies the Fokker-Planck
equation (\ref{FPE}). The initial condition (\ref{IC}) is recovered
from the Gaussian integral as $\Delta t\to0$. The boundary condition
that $p_{OUT}^{(0)}(y,t)$ satisfies can be determined only after the
boundary layer is resolved by matching. The leading order matching
condition of the boundary layer and the outer solutions is
\begin{eqnarray*}
\lim_{\eta\to\infty} p_{BL}^{(0)}(\eta,t) = p_{OUT}^{(0)}(0,t).
\end{eqnarray*}
Therefore
\begin{equation}
p_{BL}^{(0)}(\eta,t) = p_{OUT}^{(0)}(0,t).\label{pBL0-const}
\end{equation}
The matching condition at order $\sqrt{\Delta t}$ gives
 \beqq
\eta\frac{\p p^{(0)}_{OUT}(0,t)}{\p y}+p^{(1)}_{OUT}(0,t)\sim
p^{(1)}_{BL}(\eta,t)\quad\mbox{for}\quad{\eta\to\infty},
 \eeqq
which means that $p^{(1)}_{BL}(\eta,t)$ is asymptotically a linear
function of $\eta$, therefore the limit of its derivative is a
constant. Thus the matching condition reduces to
\begin{equation}
\lim_{\eta\to\infty} \frac{\p p_{BL}^{(1)}(\eta,t)}{\p \eta} =
\frac{\p p_{OUT}^{(0)}(0,t)}{\p y}.\label{matching1}
\end{equation}
The first order boundary layer term satisfies the integral equation
\begin{eqnarray}
&&p_{BL}^{(1)}(\eta,t) = \int_0^{\infty}
\frac{p_{BL}^{(1)}(\xi,t)}{\sqrt{4\pi\sigma(0,t)}}\left\{\exp\left[-\frac{(\eta-\xi)^2}{4\sigma(0,t)}
\right] + \exp\left[-\frac{(\eta+\xi)^2}{4\sigma(0,t)} \right]
\right\}\,d\xi - \label{pBL1-int} \\
&&\nonumber\\
 &&  P\int_0^{\infty}
\frac{p_{BL}^{(0)}(\xi,t)}{\sqrt{4\pi\sigma(0,t)}}\exp\left[-\frac{(\eta+\xi)^2}{4\sigma(0,t)}
\right]\,d\xi- \nonumber \\
&&\nonumber\\
&& \frac{\sigma_y(0,t)}{2\sigma(0,t)}\int_0^{\infty}
\frac{p_{BL}^{(0)}(\xi,t)}{\sqrt{4\pi\sigma(0,t)}}\,\xi\left\{\exp\left[-\frac{(\eta-\xi)^2}{4\sigma(0,t)}
\right]+\exp\left[-\frac{(\eta+\xi)^2}{4\sigma(0,t)}
\right]\right\}\,d\xi+ \nonumber \\
&&\nonumber\\
&&  \frac{\sigma_y(0,t)}{4\sigma(0,t)^2}\int_0^{\infty}
\frac{p_{BL}^{(0)}(\xi,t)}{\sqrt{4\pi\sigma(0,t)}}\,\xi\left\{(\eta-\xi)^2\exp\left[-\frac{(\eta-\xi)^2}{4\sigma(0,t)}
\right]+(\eta+\xi)^2\exp\left[-\frac{(\eta+\xi)^2}{4\sigma(0,t)}
\right]\right\}\,d\xi +\nonumber \\
&&\nonumber\\
&& \frac{2a(0,t)}{4\sigma(0,t)}\int_0^{\infty}
\frac{p_{BL}^{(0)}(\xi,t)}{\sqrt{4\pi\sigma(0,t)}}\left\{(\eta-\xi)\exp\left[-\frac{(\eta-\xi)^2}{4\sigma(0,t)}
\right]-(\eta+\xi)\exp\left[-\frac{(\eta+\xi)^2}{4\sigma(0,t)}
\right]\right\}\,d\xi.\nonumber
\end{eqnarray}
Evaluating explicitly the last four integrals in (\ref{pBL1-int})
and using (\ref{pBL0-const}), gives
\begin{eqnarray}
p_{BL}^{(1)}(\eta,t) &=& \int_0^{\infty}
\frac{p_{BL}^{(1)}(\xi,t)}{\sqrt{4\pi\sigma(0,t)}}\left\{\exp\left[-\frac{(\eta-\xi)^2}{4\sigma(0,t)}
\right] + \exp\left[-\frac{(\eta+\xi)^2}{4\sigma(0,t)} \right]
\right\}\,d\xi - \label{pBL1-int1}\\
&&\nonumber\\
&&  \frac{P}{2}\,p_{OUT}^{(0)}(0,t)\operatorname{erfc}
\left(\ds{\frac{\eta}{2\sqrt{\sigma(0,t)}}}\right) +
\frac{\sigma_y(0,t)-a(0,t)}{\sqrt{\pi\sigma(0,t)}}p_{OUT}^{(0)}(0,t)
\exp\left[-\frac{\eta^2}{4\sigma(0,t)} \right]. \nonumber
\end{eqnarray}
Differentiating (\ref{pBL1-int1}) with respect to $\eta$ and
integrating by parts, we obtain
\begin{eqnarray}
&&\frac{\p p_{BL}^{(1)}(\eta,t)}{\p \eta}
=\frac{1}{\sqrt{4\pi\sigma(0,t)}}\int_0^{\infty} \frac{\p
p_{BL}^{(1)}(\xi,t)}{\p
\eta}\left\{\exp\left[-\frac{(\eta-\xi)^2}{4\sigma(0,t)} \right] -
\exp\left[-\frac{(\eta+\xi)^2}{4\sigma(0,t)} \right] \right\}\,d\xi
+\nonumber\\
&&\nonumber\\
&&\frac{P}{2\sqrt{\pi\sigma(0,t)}}\,p_{OUT}^{(0)}(0,t)
\exp\left[\frac{-\eta^2}{4\sigma(0,t)}
\right]-\frac{\sigma_y(0,t)-a(0,t)}{2\sqrt{\pi}\,\sigma(0,t)^{3/2}}p_{OUT}^{(0)}(0,t)
\,\eta\exp\left[\frac{-\eta^2}{4\sigma(0,t)} \right].
\label{pBL1-int2}
\end{eqnarray}
Setting
\begin{equation}
g(\eta,t) = \frac{\p p_{BL}^{(1)}(\eta,t)}{\p \eta} -
\frac{P}{2\sqrt{\pi\sigma(0,t)}}\,p_{OUT}^{(0)}(0,t)
\exp\left[-\frac{\eta^2}{4\sigma(0,t)} \right],
\end{equation}
we rewrite (\ref{pBL1-int2}) as
\begin{eqnarray}
&&g(\eta,t) =  \label{pBL-int3}\\
&&\nonumber\\
 && \phi(\eta,t) +\frac{1}{\sqrt{4\pi\sigma(0,t)}}\int_0^{\infty}
g(\xi,t)\left\{\exp\left[-\frac{(\eta-\xi)^2}{4\sigma(0,t)} \right]
- \exp\left[-\frac{(\eta+\xi)^2}{4\sigma(0,t)} \right]
\right\}\,d\xi ,\nonumber
\end{eqnarray}
where
\begin{eqnarray}
\phi(\eta,t) &=&
\frac{P}{\sqrt{8\pi\sigma(0,t)}}\,p_{OUT}^{(0)}(0,t)
\exp\left[\frac{-\eta^2}{8\sigma(0,t)} \right]
\operatorname{erf}\left(\frac{\eta}{\sqrt{8\sigma(0,t)}}
\right)-\label{phiquation} \\
&&\nonumber\\
 &&
\frac{\sigma_y(0,t)-a(0,t)}{2\sqrt{\pi}\,\sigma(0,t)^{3/2}}p_{OUT}^{(0)}(0,t)
\,\eta\exp\left[\frac{-\eta^2}{4\sigma(0,t)} \right].\nonumber
\end{eqnarray}
Since $\phi(\eta,t)$ is an odd function of $\eta$, we can define
$g(\eta, t)$ for negative values as an odd function by setting
$g(\eta,t)=-g(-\eta,t)$ for $\eta<0$. Then (\ref{pBL-int3}) can be
rewritten as
\begin{equation}
g(\eta,t) = \phi(\eta,t) +
\frac{1}{\sqrt{4\pi\sigma(0,t)}}\int_{-\infty}^{\infty}
g(\xi,t)\exp\left[-\frac{(\eta-\xi)^2}{4\sigma(0,t)} \right]
\,d\xi,\label{WHE}
\end{equation}
which in Fourier space is
\begin{equation}
\hat{g}(k,t)=
\frac{\hat{\phi}(k,t)}{1-\exp[-\sigma(0,t)k^2]}.\label{gkt}
\end{equation}
Using the Wiener-Hopf method, we decompose
 \beq \hat{g}(k,t)=
\hat{g}_+(k,t) + \hat{g}_-(k,t),\label{g+-}
 \eeq
where $g_{+}(\eta)=g(\eta)\chi_{[0,\infty)}(\eta),\
g_{-}(\eta)=g(\eta)\chi_{(-\infty,0]}(\eta)$. The Fourier transform
$\hat{g}(k,t)$ exists in the sense of distributions, and
$\hat{g}_{\pm}(k,t)$ are analytic in the upper and lower halves of
the complex plane, respectively. Taylor's expansion of
$\hat{\phi}(k,t)$ in eq.(\ref{phiquation}) gives
\begin{equation}
\hat{\phi}(k,t)=
2ip_{OUT}^{(0)}(0,t)\left\{\frac{P\sqrt{\sigma(0,t)}}{\sqrt{\pi}} -
[\sigma_y(0,t)-a(0,t)]\right\}k +O(k^3)\quad \mbox{as }\quad k\to
0.\label{phihat}
\end{equation}
The the non-zero poles of (\ref{gkt}) split evenly between
$\hat{g}_+(k,t)$ and $\hat{g}_-(k,t)$, and using $\hat{g}_+(k,t)
=-\hat{g}_-(-k,t)$, the pole at the origin gives
\begin{equation}
\hat{g}_+(k,t) =
ip_{OUT}^{(0)}(0,t)\left\{\frac{P}{\sqrt{\pi\sigma(0,t)}} -
\frac{\sigma_y(0,t)-a(0,t)}{\sigma(0,t)}\right\}\frac{1}{k}+O(k)
\quad \mbox{as } \quad k\to 0.
\end{equation}
Inverting the Fourier transform $\hat{g}_+(k,t)$, by closing the
contour of integration around the lower half plane, we obtain
\begin{equation}
\lim_{\eta\to\infty}\frac{\p p_{BL}^{(1)}(\eta,t)}{\p \eta} =
p_{OUT}^{(0)}(0,t)\left\{\frac{P}{\sqrt{\pi\sigma(0,t)}} -
\frac{\sigma_y(0,t)-a(0,t)}{\sigma(0,t)}\right\}.
\end{equation}
The matching condition (\ref{matching1}) implies
\begin{equation}
\frac{\p p_{OUT}^{(0)}(0,t)}{\p y} =
p_{OUT}^{(0)}(0,t)\left\{\frac{P}{\sqrt{\pi\sigma(0,t)}} -
\frac{\sigma_y(0,t)-a(0,t)}{\sigma(0,t)}\right\}.
\end{equation}
Multiplying by $\sigma(0,t)$ and rearranging, we obtain the
radiation boundary condition
\begin{equation}
-J(0,t) = \frac{\p}{\p
y}\left[\sigma(0,t)p_{OUT}^{(0)}(0,t)\right]-a(0,t)p_{OUT}^{(0)}(0,t)
= \frac{P\sqrt{\sigma(0,t)}}{\sqrt{\pi}}p_{OUT}^{(0)}(0,t).
\end{equation}
Since $p(y,t)=p_{OUT}^{(0)}(y,t)$, the reactive ``constant"  in
(\ref{radiation}) is
\begin{equation}
\label{reactive_constant}
\kappa(t)=\frac{P\sqrt{\sigma(0,t)}}{\sqrt{\pi}}.
\end{equation}

\section{Numerical simulations in one dimension}{\label{sec:numerical}}
The explicit analytical solution of the FPE (\ref{FPE}) with the
initial condition (\ref{IC}) and the radiation boundary condition
(\ref{radiation}) for the case of vanishing drift ($a=0$) and
constant diffusion coefficient ($\sigma(x,t)=\sigma$) was first
given by Bryan in 1891 \cite{Bryan} (see \cite[$\S$14.2,
p.358]{CarslawJaeger})
\begin{eqnarray}
p(x,t\,|\,x_0) &=& \frac{1}{\sqrt{4\pi \sigma
t}}\left[\exp\left\{-\frac{(x-x_0)^2}{4\sigma t} \right\} +
\exp\left\{-\frac{(x+x_0)^2}{4\sigma t} \right\}\right] \nonumber
\\
&&\label{analytic}\\
&& -\frac{\kappa}{\sigma}\exp\left\{\frac{\kappa(x+x_0+\kappa
t)}{\sigma} \right\}\,\operatorname{erfc}\left[\frac{x+x_0+2 \kappa
t}{\sqrt{4\sigma t}} \right].\nonumber
\end{eqnarray}
The first term in (\ref{analytic}) is the fundamental solution of
(\ref{FPE}) and (\ref{IC}) with a reflecting boundary condition,
whereas the second term may be transformed into
$$-\frac{\kappa}{\sqrt{\pi \sigma^3 t}} \int_0^\infty
\exp\left\{-\frac{\kappa\xi}{\sigma}
\right\}\exp\left\{-\frac{(x+x_0+\xi)^2}{4\sigma t}\right\}\,d\xi,$$
which represents the density due to a line of exponentially
decreasing sinks extending from $-x_0$ to $-\infty$. The method of
Laplace transforming (\ref{FPE}) with respect to $t$ was later
employed \cite{LammSchultenII}, \cite{Agmon} to obtain explicit
analytical solution for the FPE (\ref{FPE})-(\ref{IC}) with
constants diffusion coefficient and (not necessarily vanishing)
drift term $a(x,t)=a$
\begin{eqnarray}
&&p(x,t\,|\,x_0) = \nonumber
\\&&\nonumber\\
&&\frac{1}{\sqrt{4\pi \sigma
t}}\left[\exp\left\{-\frac{(x-x_0-at)^2}{4\sigma t} \right\} +
\exp\left\{-\frac{ax_0}{\sigma}-\frac{(x+x_0-at)^2}{4\sigma t}
\right\}\right] \nonumber \\
&&\label{analytic2}\\
&&
-\frac{2\kappa+a}{2\sigma}\exp\left\{\frac{ax+\kappa[x+x_0+(\kappa+a)t]}{\sigma}
\right\}\,\operatorname{erfc}\left[\frac{x+x_0+(2\kappa+a)t}{\sqrt{4\sigma
t}} \right].\nonumber
\end{eqnarray}
Setting $\kappa=0$ in (\ref{analytic2}) reduces to Smoluchowski's
\cite{Gravitation} explicit analytical solution for a reflecting
boundary with a constant drift term, while setting $a=0$ reduces to
Bryan's solution (\ref{analytic}).

We conducted several numerical experiments in which $n=10^7$
trajectories were simulated according to the Euler scheme
(\ref{Euler}) with the boundary behavior (\ref{design}). The
diffusion coefficient was constant $\sigma=1$ and the reactive
constant was $\kappa=1$, giving $P=\sqrt{\pi}$ in
eq.(\ref{reactive_constant}). The trajectories were initially
located at $x_0=1$, their statistics were collected at time $t=1$,
and compared to the predicted $p(x,t=1\,|\,x_0=1)$. The convergence
of the scheme was tested by using four different time steps $\Delta
t=10^{-1},10^{-2},10^{-3},10^{-4}$.

\begin{figure}[h]
\begin{center}
{\epsfxsize=300pt\epsffile{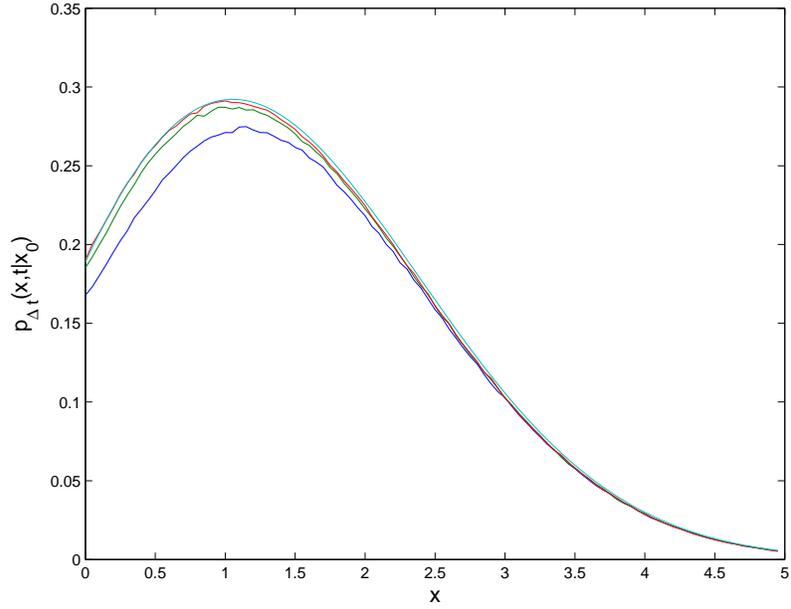}} \caption{No drift:
The analytical solution (\ref{analytic}) (Magenta), and the three
numerical densities $\Delta t=10^{-1}$ (Blue), $\Delta t=10^{-2}$
(Green), $\Delta t=10^{-3}$ (Red) approaching it from below. The
numerical density of $\Delta t = 10^{-4}$ is not shown, because it
is difficult to distinguish it from the analytic density.
(Parameters: $\sigma=\kappa=x_0=t=1$, $a=0$, $P=\sqrt{\pi}$,
$n=10^7$)}\label{f:no_drift}
\end{center}
\end{figure}

The first experiment corresponds to a vanishing drift $a=0$. Figure
\ref{f:no_drift} shows the convergence of the numerical scheme to
the analytic solution (\ref{analytic}). The rate of convergence of
the numerical scheme to the analytic solution is $\sqrt{\Delta t}$.
This is demonstrated, for example, by the survival probability
$$p_{sur}(x_0,t)=\int_0^{\infty} p(x,t\,|\,x_0)\,dx$$ of finding the
trajectory inside the domain at time $t$, that is, the probability
that the trajectory was not absorbed prior to $t$. Integrating
(\ref{analytic}) gives $p_{sur}(1,1)=0.77095\ldots$ for
$\sigma=\kappa=1$. The survival probability is estimated numerically
by the ratio of the number of survived (unabsorbed) trajectories
$n_{sur}$ and the total number of simulated trajectories $n=10^7$.
Table \ref{t:surv} shows that the convergence rate of the estimated
survival probability to its analytic value is $\sqrt{\Delta t}$ as
predicted by our boundary layer analysis. The statistical estimation
(variance) error due to the finite number of simulated trajectories
is $\sqrt{p_{sur}(1-p_{sur})/n}=0.00013\ldots$, which is an order of
magnitude smaller than the smallest (bias) error obtained for
$\Delta t=10^{-4}$ (see Table \ref{t:surv}).

%
%\begin{figure}[h]
%\begin{center}
%{\epsfxsize=300pt\epsffile{radiation_sqrt.eps}} \caption{Log scale
%plot of the survival probability estimation error against $\Delta
%t$. The linear fit gives a slope $\approx 0.5$ that corresponds to a
%convergence rate of $\sqrt{\Delta t}$. (Parameters:
%$\sigma=\kappa=x_0=t=1$, $a=0$, $P=\sqrt{\pi}$,
%$n=10^7$)}\label{f:surv}
%\end{center}
%\end{figure}

\begin{figure}[h]
\begin{center}
{\epsfxsize=300pt\epsffile{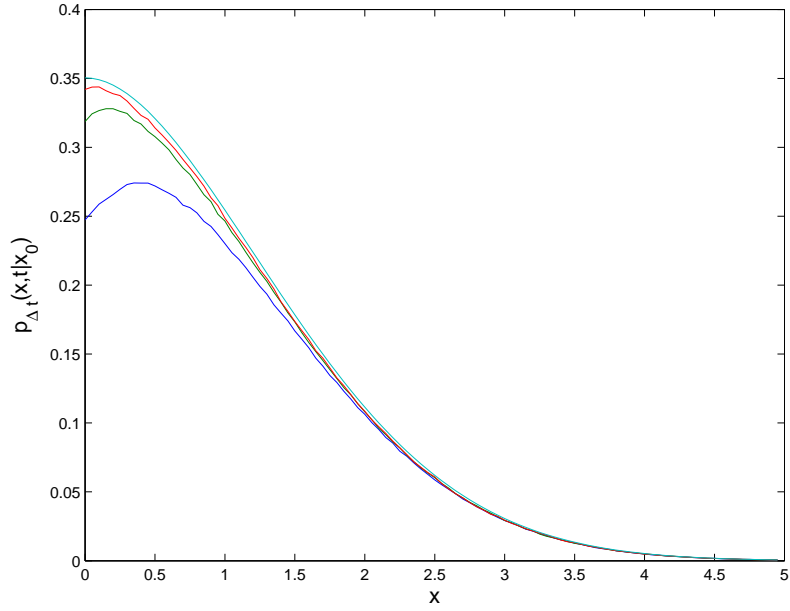}} \caption{Drift,
$a=-1$: The analytical solution (\ref{analytic2}) (Magenta) and the
numerical densities $\Delta t=10^{-1}$ (Blue), $\Delta t=10^{-2}$
(Green), $\Delta t=10^{-3}$ (Red) that approach it from below.
(Parameters: $\sigma=\kappa=x_0=t=1$, $P=\sqrt{\pi}$,
$n=10^7$)}\label{f:drift}
\end{center}
\end{figure}

In the second experiment the drift term $a=-1$ shifts the density
leftwards, and causes more trajectories to react with the boundary.
Figure \ref{f:drift} shows the convergence of the numerical scheme
to the analytic solution (\ref{analytic2}).

\begin{figure}[h]
\begin{center}
{\epsfxsize=300pt\epsffile{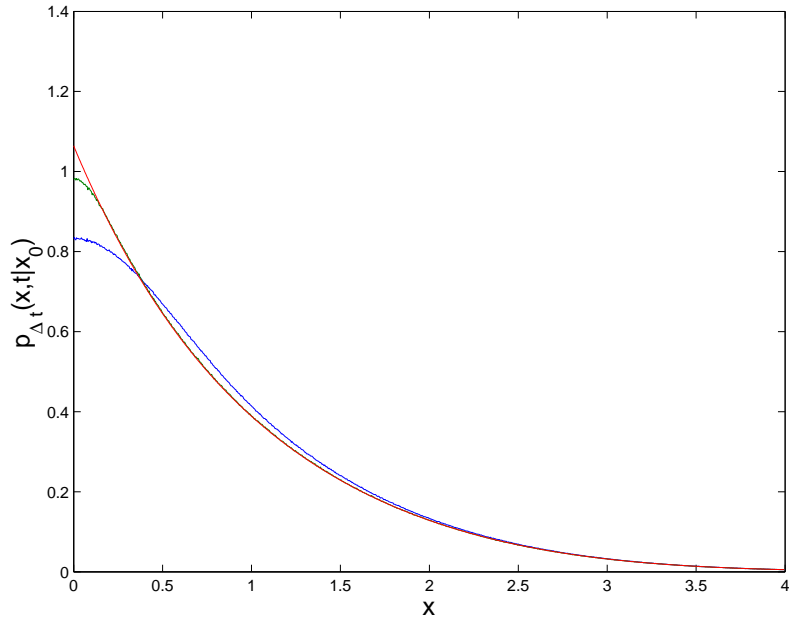}} \caption{Drift,
$a=-1$, reflecting boundary $P=\kappa=0$: The analytic solution
(\ref{analytic2}) (Red) and the numerical densities $\Delta
t=10^{-1}$ (Blue) and $\Delta t= 10^{-2}$ (Green) with $n=10^8$
simulated trajectories to obtain a finer boundary resolution.
(Parameters: $\sigma=\kappa=x_0=t=1$)}\label{f:reflecting}
\end{center}
\end{figure}

The final experiment corresponds to a reflecting boundary,
$P=\kappa=0$ and a constant non-vanishing drift towards the boundary
$a=-1$. We simulated $n=10^8$ trajectories to obtain a finer
resolution at the boundary. Figure \ref{f:reflecting} shows a
comparison between the analytical solution (\ref{analytic2}) and the
numerical densities for $\Delta t=10^{-1},10^{-2}$. The no flux
condition $J=0$ of a reflecting boundary together with (\ref{J})
gives a negative boundary derivative $p_y(0,t) = -p(0,t) < 0$. In
particular, the analytic solution (\ref{analytic2}) satisfies
$p_y(0,1) =-p(0,1)= -(2+\sqrt{\pi})/(2\sqrt{\pi})\approx -1.06$. The
numerical densities, however, are flat at the boundary. Their first
derivatives vanish at the boundary, as predicted in (\ref{px0}) and
shown in Figure \ref{f:reflecting}. The first derivative changes
from $0$ to $O(1)$ on an interval of length $O(\sqrt{\Delta t})$,
manifesting a boundary layer behavior, though there is no such
behavior in the density itself.

\section{Diffusion in  $\rR^d$ with partial
oblique reflection at the boundary} We consider the $d$-dimensional
stochastic dynamics
 \beq \label{SDEn1}
 \dot{\x}=\mb{a}(\x,t)+ \sqrt{2}\mb{B}(t) \, \dot{\w}
 \eeq
in the half space
$$\Omega=\{\x=(x_1,x_2,\ldots,x_d)\in \rR^d : x_1 > 0 \}$$
where $\w$ is a vector of $d$ independent Brownian motions and we
assume that the diffusion tensor
$\mb{\sigma}(t)=\mb{B}(t)\mb{B}^T(t)$ is uniformly positive definite
for all $t\geq s$. The case of space-dependent diffusion involves
many technically complicated calculations and will be considered in
a separate paper. We use henceforth the abbreviation
$\mb{\sigma}(t)=\mb{\sigma}$. The radiation condition
(\ref{radiation}) becomes
 \beq
-\J(\y,t\,|\,\x,s)\cdot\mb{n}=\kappa(\y,t)p(\y,t\,|\,\x,s),\quad\mbox{for}\quad\y\in\p\Omega,\
\x\in\Omega,\label{radiation-n1}
 \eeq
where the components of the flux vector $\J(\y,t\,|\,\x,s)$ are
defined by
 \beq
J^k(\y,t\,|\,\x,s)=-[a^k(\y,t)p(\y,t\,|\,\x,s)]+\sum_{j=1}^d
\frac{\p}{\p
y_j}\left[\sigma^{j,k}p(\y,t\,|\,x,s)\right],\label{Fluxn1}
 \eeq
where $\sigma^{j,k}$ are the elements of the diffusion matrix $\mb{\sigma}$.
The Fokker-Plank equation for the pdf of $\x(t)$ can be written as
 \beq
\frac{\p p(\y,t\,|\,\x,s)}{\p t}
=-\nabla_{\y}\cdot\J(\y,t\,|\,\x,s)\quad\mbox{for all} \quad \y,\x
\in \Omega.\label{FPEd}
 \eeq

If $\x\in\Omega$, but
$$\x'=\x+\A(\x,t)\Delta t+\sqrt{2}\B(t)\,\Delta \w(t,\Delta
 t)\not\in\Omega,$$
the Euler scheme for (\ref{SDEn1}) with oblique reflection in
$\p\Omega$ reflects the point $\x'$ obliquely in the constant
direction of $\U$ to a point $\x''\in\Omega$, as described below.
First, we denote by $\x'_B$ the normal projection of a point $\x'$
on $\p\Omega$, that is, $\x'_B=\x'-(\x'\cdot\n)\n$. Then we write
the Euler scheme for (\ref{SDEn1}) with partially reflecting
boundary as
 \beq
\x(t+\Delta t)=\left\{\begin{array}{l}\x'\quad\mbox{for}\quad\x'\in\Omega\\
\\
\x''\quad\mbox{w.p.}\quad 1-P\left({\x'_B}\right)\sqrt{\Delta
t},\quad\mbox{if}\quad\x'\not\in\Omega,\\
\\
\mbox{terminate trajectory w.p.}\ P\left({\x'_B}\right)\sqrt{\Delta
t},\quad\mbox{if}\quad\x'\not\in\Omega.
\end{array}\right.\label{Euler-n1}
 \eeq
The value of the termination probability
$P\left({\x'_B}\right)\sqrt{\Delta t}$, that varies continuously in
the boundary, is evaluated at the normal projection of the point
$\x'$ on the boundary. The oblique reflection in the direction of
the unit vector $\U$ ($v_1\neq0$) is defined by
\begin{equation}
\x'' = \x' - \frac{2x'_1}{v_1}\U.
\label{Euler-n2}
\end{equation}
Note that $x''_1=-x'_1$ guarantees that the reflected point of a
crossing trajectory is inside the domain $\Omega$. The fact that the
normal components of $\x''$ and $\x'$ are of equal lengths makes the
high-dimensional boundary layer analysis similar to that in one
dimension. Normal reflection corresponds to
$\mb{v}=\mb{n}=(1,0,\ldots,0)$.

\begin{figure}[h]
\begin{center}
{\epsfxsize=300pt\epsffile{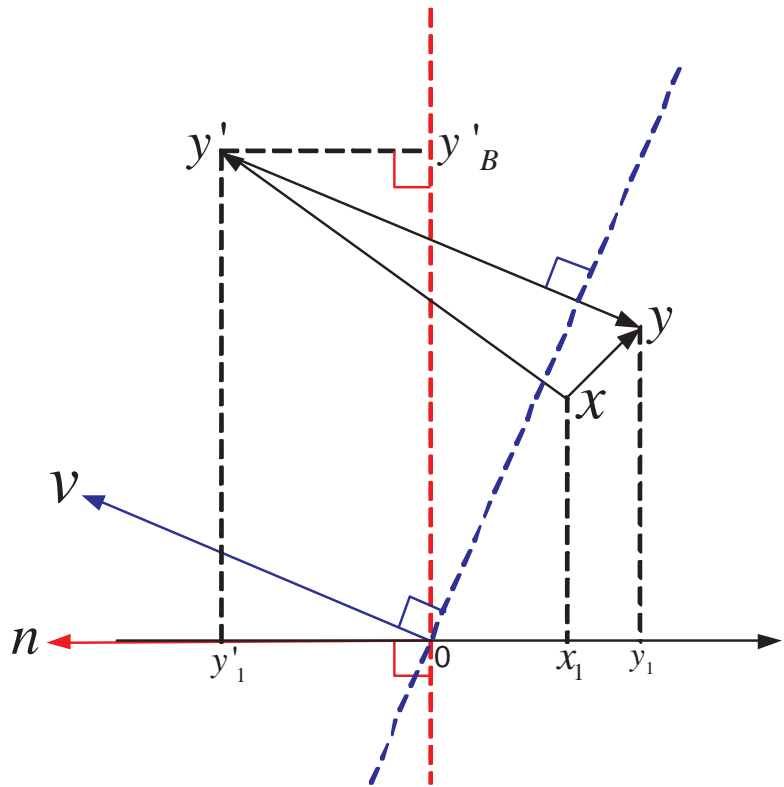}} \caption{A simulated
trajectory can get from $\x$ to $\y$ in a single time step $\Delta
t$ in two different ways: (i) directly from $\x$ to $\y$, without
crossing the boundary, and (ii) by crossing the boundary from $\x$
to $\y'$ and reflection in the oblique direction $\U$ with
probability $1-P(\y'_B)\sqrt{\Delta t}$ to $\y$. The reflection law
(\ref{Euler-n1})-(\ref{yy'}) satisfies $y_1' =
-y_1$.}\label{f:reflection}
\end{center}
\end{figure}

We note that for a point $\y\in\Omega$, we can write
$\Pr\{\x''=\y\}=\Pr\{\x'=\y'\}$, where
 \beq
\y = {\y}' -\frac{2\y'\cdot\n}{v_1}\U\label{yy'}
 \eeq
is the oblique reflection of $\y'$ (see fig. \ref{f:reflection}).
Given $\y$, equation (\ref{yy'}) defines $\y'$ as
 \beq
 \y'=\y-2\frac{y_1}{v_1}\U.
 \eeq

As in the one-dimensional case, the forward Kolmogorov equation is
\begin{eqnarray}
p_{\Delta t}(\mb{y},t+\Delta t) &=& \int_{x_1>0} \frac{p_{\Delta
t}(\x,t)}{(4\pi\Delta t)^{d/2}\sqrt{\det \mb{\sigma}}}
\Bigg\{\exp\left[-\frac{{\cal{B}}(\mb{x}+\mb{a}(\x,t)\Delta
t,\mb{y})}{4\Delta t} \right] + \nonumber\\
&&\nonumber\\
&&  (1-P(\y'_B)\sqrt{\Delta
t})\exp\left[-\frac{{\cal{B}}(\mb{x}+\mb{a}(\x,t)\Delta
t,\mb{y}')}{4\Delta t} \right] \Bigg\}\,d\x,\label{d-prop}
\end{eqnarray}
where
\begin{equation}
{\cal{B}}(\x, \y) = (\x-\y)^T \mb{\sigma}^{-1} (\x-\y).
\end{equation}
We construct a boundary layer of width $O(\sqrt{\Delta t})$ in the
normal direction to the boundary. The layer extends infinitely in
the $d-1$ directions tangent to the boundary
\begin{equation}
p_{BL}(\eta_1,y_2,\ldots,y_d,t) = p_{\Delta t}(\eta_1\sqrt{\Delta
t}, y_2,\ldots,y_d,t).
\end{equation}
In other words, $p_{BL}(\eta_1 \mb{n} + \mb{y}_B,t) = p_{\Delta
t}(\eta_1\sqrt{\Delta t}\,\mb{n} + \mb{y}_B,t)$, where $\mb{y}_B =
(0,y_2,y_3,\ldots,y_d)$. As in the one-dimensional case, we assume
the asymptotic expansion
\begin{eqnarray}
p_{BL}(\eta_1\mb{n}+\mb{y}_B,t) &\sim&
p_{BL}^{(0)}(\eta_1\mb{n}+\mb{y}_B,t) + \sqrt{\Delta t} \,
p_{BL}^{(1)}(\eta_1\mb{n}+\mb{y}_B,t) + \ldots.\label{pblexp}
\end{eqnarray}
and substitute
\begin{equation}
\mb{x} = \mb{y}_B + \sqrt{\Delta t}\,\mb{\xi}
\end{equation}
in the integral (\ref{d-prop}). We obtain
\begin{eqnarray}
&&p_{BL}(\eta_1\mb{n}+\mb{y}_B,t+\Delta t)= \int_{\xi_1>0}
\frac{p_{BL}(\xi_1\mb{n}+\mb{y}_B+\sqrt{\Delta
t}\,\mb{\xi}_B,t)}{(4\pi)^{d/2}\sqrt{\det \mb{\sigma}}}
\times\label{I+II}
\\
&&\nonumber\\
 && \left\{\exp\left[-\frac{{\cal{B}}(\mb{\xi}+\mb{a}(\y_B,t)
 \sqrt{\Delta t},\eta_1
\mb{n})}{4} \right] + (1-P(\mb{y}'_B )\sqrt{\Delta
t})\times\right.\nonumber\\
&&\nonumber\\
&&\left.\exp\left[-\frac14{\cal{B}}\left(\mb{\xi}+\mb{a}(\y_B,t)
 \sqrt{\Delta t},\eta_1\mb{n}-\ds{\frac{2\eta_1}{v_1}}\mb{v}\right)
\right] \right\}\,d\mb{\xi}+O(\Delta t).\nonumber
\end{eqnarray}
We calculate separately the integral of the first and second terms
in the braces. Substituting in the first integral
 \begin{eqnarray}
\mb{z} &=& \mb{\sigma}^{-1/2}(\mb{\xi}-\eta_1\mb{n}),
\end{eqnarray}
transforms the domain of integration to
\begin{equation}
\mb{z}\cdot \tilde{\mb{n}} > -\frac{\eta_1}{\sqrt{\sigma_n}},
\end{equation}
where $\tilde{\mb{n}} =
\frac{\mb{\sigma}^{1/2}\mb{n}}{\|\mb{\sigma}^{1/2}\mb{n} \|}$ is a
unit vector, and $\sigma_n = \mb{n}^T \mb{\sigma} \mb{n} =
\|\mb{\sigma}^{1/2}\mb{n} \|^2$. Similarly, we transform the second
integral by substituting $\mb{z}' =
\mb{\sigma}^{-1/2}\left(\mb{\xi}-\eta_1\mb{n}+
\ds{\frac{2\eta_1}{v_1}}\,\mb{v}\right)$. Using the expansion
(\ref{pblexp}), we obtain at the leading order the integral equation
\begin{eqnarray*}
&& p_{BL}^{(0)}(\eta_1\mb{n}+\mb{y}_B,t) = \\
&&\\
 && \frac{1}{(4\pi)^{d/2}} \int_{\mb{z}\cdot \tilde{\mb{n}} >
-\frac{\eta_1}{\sqrt{\sigma_n}}}
p_{BL}^{(0)}\left((\eta_1+\sqrt{\sigma_n}\,\mb{z}\cdot
\tilde{\mb{n}})\mb{n}+\mb{y}_B,t\right)\exp\left[-\frac{\|
\mb{z}\|^2}{4} \right]\,d\mb{z} + \nonumber \\
&&\\
&& \frac{1}{(4\pi)^{d/2}} \int_{\mb{z}'\cdot \tilde{\mb{n}}
> \frac{\eta_1}{\sqrt{\sigma_n}}}
p_{BL}^{(0)}\left((-\eta_1+\sqrt{\sigma_n}\,\mb{z}'\cdot\tilde{\mb{n}})\mb{n}+
\mb{y}_B,t\right)\exp\left[-\frac{\| \mb{z}'\|^2}{4}
\right]\,d\mb{z}'.\nonumber
\end{eqnarray*}
Integrating in the $d-1$ directions orthogonal to $\tilde{\mb{n}}$,
yields
\begin{eqnarray*}
&&p_{BL}^{(0)}(\eta_1\mb{n}+\mb{y}_B,t) = \frac{1}{\sqrt{4\pi}}
\int_{-\frac{\eta_1}{\sqrt{\sigma_n}}}^\infty
p_{BL}^{(0)}\left((\eta_1+\sqrt{\sigma_n}\,u)\mb{n}+
\mb{y}_B,t\right)\exp\left[-\frac{u^2}{4} \right]\,du + \nonumber \\
&&\\
 && \frac{1}{\sqrt{4\pi}}
\int_{\frac{\eta_1}{\sqrt{\sigma_n}}}^\infty
p_{BL}^{(0)}\left((-\eta_1+\sqrt{\sigma_n}\,u)\mb{n}+\mb{y}_B,t\right)
\exp\left[-\frac{u^2}{4}
\right]\,du=\nonumber \\
&&\nonumber\\
 && \frac{1}{\sqrt{4\pi\sigma_n}} \int_{0}^\infty
p_{BL}^{(0)}\left(u\mb{n}+\mb{y}_B,t\right)\left\{\exp\left[-\frac{(u-\eta_1)^2}{4\sigma_n}
\right] + \exp\left[-\frac{(u+\eta_1)^2}{4\sigma_n} \right]
\right\}\,du. \nonumber
\end{eqnarray*}
This is the same leading order integral equation as that of the
one-dimensional case (\ref{pBL0-int}), so the solution is
independent of $\eta_1$, and matching to the outer solution gives
\begin{equation}
p_{BL}^{(0)}(\eta_1\mb{n}+\mb{y}_B,t) =
p_{OUT}^{(0)}(\mb{y}_B,t).\label{d-dim-match}
\end{equation}
To evaluate the  $O(\sqrt{\Delta t})$ terms, we expand in the first
integral in (\ref{I+II})
\begin{eqnarray}
\mathcal{B}(\mb{\xi}+\mb{a}(\y_B,t)\sqrt{\Delta t},\eta_1\mb{n}) &=&
(\mb{\xi}-\eta_1\mb{n})\cdot \mb{\sigma}^{-1}
(\mb{\xi}-\eta_1\mb{n}) + \nonumber\\
&&\nonumber\\
&&\sqrt{\Delta t}\, 2\mb{a}(\y_B,t)\cdot \mb{\sigma}^{-1}
(\mb{\xi}-\eta_1\mb{n}).
\end{eqnarray}
and in the second integral
 \beq
 {\cal{B}}\left(\mb{\xi}+\mb{a}(\y_B,t)
 \sqrt{\Delta
 t},\eta_1\mb{n}-\ds{\frac{2\eta_1}{v_1}}\mb{v}\right)&=&
\left(\mb{\xi}-\eta_1\mb{n}+\ds{\frac{2\eta_1}{v_1}}\mb{v}\right)\cdot
\mb{\sigma}^{-1}\left
(\mb{\xi}-\eta_1\mb{n}\ds{\frac{2\eta_1}{v_1}}\mb{v}\right)+
\nonumber\\
&&\nonumber\\
&& \sqrt{\Delta t}\, 2\mb{a}(\y_B,t)\cdot \mb{\sigma}^{-1}
\left(\mb{\xi}-\eta_1\mb{n}\ds{\frac{2\eta_1}{v_1}}\mb{v}\right).
 \eeq
The $O(\sqrt{\Delta t})$ contribution of the drift term for the
first exponential term is
\begin{eqnarray}
&& -\frac{1}{4}\int_{\xi_1>0}
\frac{p_{OUT}^{(0)}(\mb{y}_B,t)}{(4\pi)^{d/2}\sqrt{\det
\mb{\sigma}}} \exp
\left\{-\frac{\mathcal{B}(\mb{\xi},\eta_1\mb{n})}{4}
\right\}\left[2\mb{a}(\y_B,t)\cdot \mb{\sigma}^{-1}(\mb{\xi}-\eta_1
\mb{n})
\right]\,d\mb{\xi}=\nonumber \\
&&\nonumber\\
 && -\frac{1}{4}
\frac{p_{OUT}^{(0)}(\mb{y}_B,t)}{\sqrt{4\pi}} 2\mb{a}(\y_B,t)\cdot
\mb{\sigma}^{-1/2}\tilde{\mb{n}}
\int_{-\eta_1/\sqrt{\sigma_n}}^\infty ue^{-u^2/4}\,du =\nonumber \\
&&\nonumber\\
 && -\frac{1}{2}
\frac{p_{OUT}^{(0)}(\mb{y}_B,t)}{\sqrt{\pi \sigma_n}}
\mb{a}(\y_B,t)\cdot \mb{n} \exp
\left\{\frac{-\eta_1^2}{4\sigma_n}\right\}.
\end{eqnarray}
The second exponential has the same contribution, so the overall
contribution of the drift to the $O(\sqrt{\Delta t})$ term is
\begin{eqnarray}
-\frac{p_{OUT}^{(0)}(\mb{y}_B,t)}{\sqrt{\pi \sigma_n}}
\mb{a}(\y_B,t)\cdot \mb{n} \exp
\left\{\frac{-\eta_1^2}{4\sigma_n}\right\}.
\end{eqnarray}

Now, we expand
\begin{eqnarray}
&&p_{BL}^{(0)}\left((\eta_1+\sqrt{\sigma_n}\,\mb{z}\cdot
\tilde{\mb{n}})\mb{n}+\mb{y}_B + \sqrt{\Delta
t}\,(\mb{\sigma}^{1/2}\mb{z})_B,t\right) =
p_{BL}^{(0)}\left((\eta_1+\sqrt{\sigma_n}\,\mb{z}\cdot
\tilde{\mb{n}})\mb{n}+\mb{y}_B,t\right) +\nonumber \\
&&\nonumber\\
 &&  \sqrt{\Delta t}\,\nabla
p_{BL}^{(0)}\left((\eta_1+\sqrt{\sigma_n}\,\mb{z}\cdot
\tilde{\mb{n}})\mb{n}+\mb{y}_B,t\right) \cdot
(\mb{\sigma}^{1/2}\mb{z})_B + O(\Delta t).\label{expDeltat}
\end{eqnarray}
Together with (\ref{d-dim-match}), the expansion (\ref{expDeltat})
reduces to
\begin{eqnarray*}
&&p_{BL}^{(0)}\left((\eta_1+\sqrt{\sigma_n}\,\mb{z}\cdot
\tilde{\mb{n}})\mb{n}+\mb{y}_B + \sqrt{\Delta
t}\,(\mb{\sigma}^{1/2}\mb{z})_B,t\right) = \\
&&\\
 &&p_{OUT}^{(0)}\left(\mb{y}_B,t\right) + \sqrt{\Delta
t}\,\nabla p_{OUT}^{(0)}\left(\mb{y}_B,t\right) \cdot
(\mb{\sigma}^{1/2}\mb{z})_B + O(\Delta t).\nonumber
\end{eqnarray*}
Integrating as above, we obtain the $O(\sqrt{\Delta t})$ integral
equation as
\begin{eqnarray*}
&&p_{BL}^{(1)}(\eta_1\mb{n}+\mb{y}_B,t)  =\\
&&\\
&& \frac{1}{\sqrt{4\pi\sigma_n}} \int_{0}^\infty
p_{BL}^{(1)}\left(u\mb{n}+\mb{y}_B,t\right)\left\{\exp\left[-\frac{(u-\eta_1)^2}{4\sigma_n}
\right] + \exp\left[-\frac{(u+\eta_1)^2}{4\sigma_n} \right]
\right\}\,du- \nonumber \\
&&\\
 &&
\frac{P\left(\mb{y}'_B\right)p_{OUT}^{(0)}\left(\mb{y}_B,t\right)}{\sqrt{4\pi\sigma_n}}
\int_{0}^\infty \exp\left[-\frac{(u+\eta_1)^2}{4\sigma_n} \right]
\,du +\nonumber \\
&&\\
 &&
\frac{1}{\sqrt{4\pi}}\int_{\frac{\eta_1}{\sqrt{\sigma_n}}}^\infty
\nabla p_{OUT}^{(0)}(\mb{y}_B,t) \cdot
\left(2\mb{\sigma}^{1/2}u\tilde{\mb{n}}-\frac{2\eta_1}{v_1}\mb{v}
\right)_B \exp\left[-\frac{u^2}{4} \right]\,du -\nonumber \\
&&\\
&& \frac{p_{OUT}^{(0)}(\mb{y}_B,t)}{\sqrt{\pi \sigma_n}}
\mb{a}(\y_B,t)\cdot \mb{n} \exp
\left\{\frac{-\eta_1^2}{4\sigma_n}\right\}.
\end{eqnarray*}
Differentiating with respect to $\eta_1$ and integrating by parts
(as was done in the one dimensional case), we arrive at the integral
equation
\begin{eqnarray*}
&&\frac{\p p_{BL}^{(1)}(\eta_1\mb{n}+\mb{y}_B,t)}{\p n}=\\
&&\\
&& \frac{1}{\sqrt{4\pi\sigma_n}} \int_{0}^\infty \frac{\p
p_{BL}^{(1)}\left(u\mb{n}+\mb{y}_B,t\right)}{\p
n}\left\{\exp\left[-\frac{(u-\eta_1)^2}{4\sigma_n} \right] -
\exp\left[-\frac{(u+\eta_1)^2}{4\sigma_n} \right] \right\}\,du-  \\
&&\\
&&\frac{P\left(\mb{y}'_B\right)p_{OUT}^{(0)}\left(\mb{y}_B,t\right)}{\sqrt{4\pi\sigma_n}}
\exp\left[\frac{-\eta_1^2}{4\sigma_n} \right]+ \\
&&\nonumber\\
 &&  \nabla p_{OUT}^{(0)}(\mb{y}_B,t) \cdot
\left\{-\frac{1}{\sqrt{\pi\sigma_n}}\left[\frac{\mb{\sigma}\mb{n}}{\sigma_n}
-\frac{\mb{v}}{v_1}\right]
\eta_1\exp\left[\frac{-\eta_1^2}{4\sigma_n}
 \right] - \mb{v}
 \frac{\operatorname{erfc}\left(\ds{\frac{\eta_1}{2\sqrt{\sigma_n}}}\right)}{v_1}
 \right\}_B+\nonumber \\
 &&\\
&& \frac{p_{OUT}^{(0)}(\mb{y}_B,t)}{\sqrt{\pi \sigma_n}}
\mb{a}(\y_B,t)\cdot \mb{n} \frac{\eta_1}{2\sigma_n}\exp
\left[\frac{-\eta_1^2}{4\sigma_n}\right].
\end{eqnarray*}
The Wiener-Hopf method requires the extension of the
$\operatorname{erfc}$ function discontinuously as an odd function,
that is, to define
$\widetilde{\operatorname{erfc}}(x)=\operatorname{sgn}(x)\operatorname{erfc}(|x|)$.
Following the calculations of the one dimensional case, it remains
to determine the small $k$ behavior of the Fourier transform of
$\widetilde{\operatorname{erfc}}(x)$. Using
\begin{equation}
\int_{-\infty}^\infty
\widetilde{\operatorname{erfc}}\left(\frac{\eta}{2\sqrt{\sigma_n}}
\right)\exp\{ik\eta\}\,d\eta \sim 2ik \int_0^\infty
\operatorname{erfc}\left(\frac{\eta}{2\sqrt{\sigma_n}} \right) \eta
\,d\eta = 2ik\sigma_n,
\end{equation}
we obtain, as in (\ref{phihat}),
 \beqq
\hat{\phi}(k)& \sim&2ik
\Bigg{\{}\frac{P\left(\mb{y}'_B\right)p_{OUT}^{(0)}\left(\mb{y}_B,t\right)\sqrt{\sigma_n}}{\sqrt{\pi}}-
2\sigma_n\nabla p_{OUT}^{(0)}(\mb{y}_B,t) \cdot
\left[\frac{\mb{\sigma}\mb{n}}{\sigma_n}-\frac{\mb{v}}{2v_1}
\right]_B +\\
&&\\
&&p_{OUT}^{(0)}(\mb{y}_B,t) \mb{a}(\y_B,t)\cdot \mb{n}\Bigg{\}}
\quad \mbox{as} \quad k\to 0.
 \eeqq
Therefore,
 \beqq
&&\lim_{\eta_1\to\infty}\frac{\p
p_{BL}^{(1)}(\eta_1\mb{n}+\mb{y}_B,t)}{\p n} =\\
&&\\
&&
\left\{\frac{P\left(\mb{y}'_B\right)p_{OUT}^{(0)}\left(\mb{y}_B,t\right)}{\sqrt{\pi\sigma_n}}-
2\nabla p_{OUT}^{(0)}(\mb{y}_B,t) \cdot
\left[\frac{\mb{\sigma}\mb{n}}{\sigma_n}-\frac{\mb{v}}{2v_1}
\right]_B +  p_{OUT}^{(0)}(\mb{y}_B,t) \frac{\mb{a}(\y_B,t)\cdot
\mb{n}}{\sigma_n}\right\}.
 \eeqq
Combining with the matching condition
\begin{equation}
\lim_{\eta\to\infty}\frac{\p
p_{BL}^{(1)}(\eta_1\mb{n}+\mb{y}_B,t)}{\p n} = \frac{\p
p_{OUT}^{(0)}(\mb{y}_B,t)}{\p n},
\end{equation}
we obtain
 \beqq
&&\frac{\p p_{OUT}^{(0)}(\mb{y}_B,t)}{\p n} =\\
&&\\
&&
\left\{\frac{P\left(\mb{y}_B\right)p_{OUT}^{(0)}\left(\mb{y}_B,t\right)}{\sqrt{\pi\sigma_n}}-
2\nabla p_{OUT}^{(0)}(\mb{y}_B,t) \cdot
\left[\frac{\mb{\sigma}\mb{n}}{\sigma_n}-\frac{\mb{v}}{2v_1}
\right]_B + p_{OUT}^{(0)}(\mb{y}_B,t) \frac{\mb{a}(\y_B,t)\cdot
\mb{n}}{\sigma_n} \right\}.
 \eeqq
The requirement that the pdf of the limiting diffusion process
satisfies the Robin boundary condition leads to the only possible
choice
\begin{equation}
\label{co-normal}
\mb{v} = \frac{\mb{\sigma}\mb{n}}{\|\mb{\sigma}\mb{n} \|}.
\end{equation}
Otherwise, we obtain an oblique derivative boundary condition. Since
$\y'_B\to\y_B$ as $\Delta t\to0$, we obtain the Robin boundary
condition
 \beqq
-\mb{J}_{OUT}(\mb{y}_B,t)\cdot \mb{n} &=& \nabla
p_{OUT}^{(0)}(\mb{y}_B,t) \cdot \mb{\sigma}\mb{n}  -
p_{OUT}^{(0)}(\mb{y}_B,t) \mb{a}(\y_B,t)\cdot \mb{n}\\
&&\\
&=&
\frac{P\left(\mb{y}_B\right)p_{OUT}^{(0)}\left(\mb{y}_B,t\right)\sqrt{\sigma_n}}{\sqrt{\pi}}.
 \eeqq
The reflection direction $\mb{v}$ of crossing trajectories is the
co-normal direction $\mb{\sigma}\mb{n}$. Normal reflection (i.e.,
replacing $\mb{v}$ by $\mb{n}$) gives rise to the boundary normal
flux if and only if $\mb{n}$ is an eigenvector of the diffusion
tensor $\mb{\sigma}$. The limit of the outer solution as $\Delta
t\to 0$ is the solution of the Fokker-Planck equation (\ref{FPEd})
with the radiation boundary condition
 \beq
-\mb{J}(\y,t)\cdot \mb{n} =
\kappa(\y)p\left(\y,t\right)\quad\mbox{for}\quad\y\in\p\Omega,
\label{RBC-nd}
 \eeq
 where the reactive "constant" is
 \beq
\kappa(\y)=\frac{P(\y)\sqrt{\sigma_n}}{\sqrt{\pi}}.\label{reactive_constantd1}
 \eeq
Note that normal reflection will not recover the normal flux of the
radiation condition if $\n$ is not an eigenvector of $\mb{\sigma}$.

\section{Numerical simulations in two dimensions}
To illustrate the co-normal reflection law (\ref{co-normal}) in the
Euler scheme (\ref{Euler-n1})-(\ref{yy'}) in the half plane
$x\geq0$, we ran several numerical experiments. The simulations show
the convergence of the pdf of the numerical solution to that of the
FPE with the radiation boundary condition
(\ref{RBC-nd})-(\ref{reactive_constantd1}). Unlike the
one-dimensional case, no explicit solution of the anisotropic Robin
problem for the FPE in the half plane is available, so we compare
the statistics of the simulated trajectories with a numerical
solution of the FPE. The latter is constructed by the stable
Crank-Nicolson scheme on lattice points, where in each time step the
sparse linear system is solved by the conjugate gradient method.

In all numerical experiments the initial point is $(x_0, y_0) =
(0.3, 0)$ and the statistics is collected at time $T = 0.5$. We
choose the reactive constant $\kappa=1$ and the diffusion matrix
$\mb{B}$ in (\ref{SDEn1})
 \beqq
\B =\begin{pmatrix} 0.3 & 0.4 \\ 0 & 1 \end{pmatrix},
 \eeqq
which gives the anisotropic diffusion tensor
 \beqq
\mb{\sigma} = \mb{B}\mb{B}^T =
\begin{pmatrix} 0.25 & 0.4 \\ 0.4 & 1 \end{pmatrix}.
 \eeqq
We simulate $n = 10^7$ trajectories with time steps $\Delta t =
10^{-1}, 10^{-2}, 10^{-3}, 10^{-4}$ in each experiment.

\begin{figure} [htbp]
\begin{center}
{\epsfxsize=400pt\epsffile{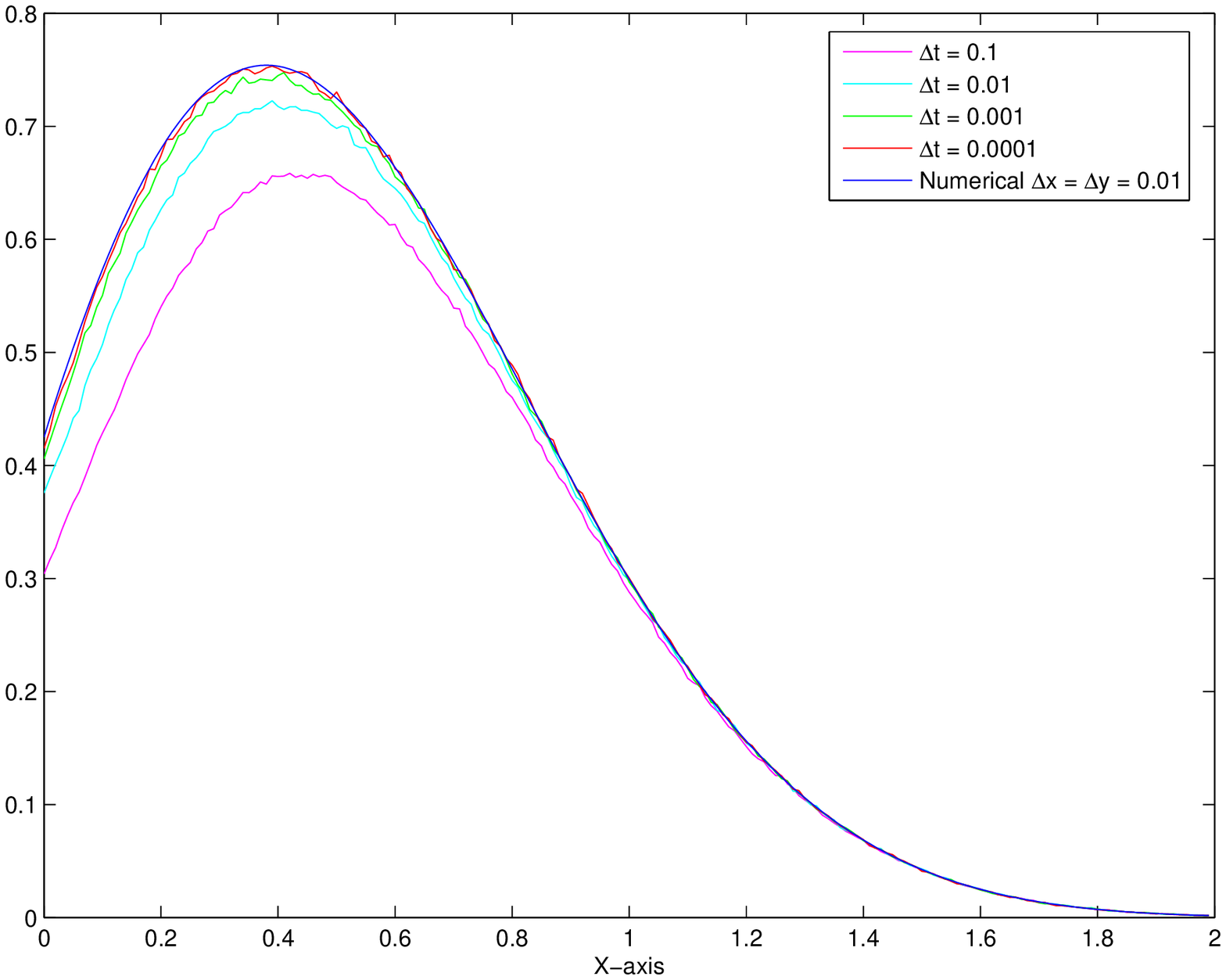}} \caption {The
marginal density of $x(T)$ with no drift and correct oblique
reflection (the first experiment). The numerical solution of the FPE
(blue) with grid size $\Delta x = 0.01$ and estimates from the
simulation of $n = 10^7$ trajectories with time steps $\Delta t =
10^{-1}, 10^{-2}, 10^{-3}, 10^{-4}$. }
\label{fig:implicit_correct_x}
\end{center}
\end{figure}

\begin{figure} [htbp]
\begin{center}
{\epsfxsize=400pt\epsffile{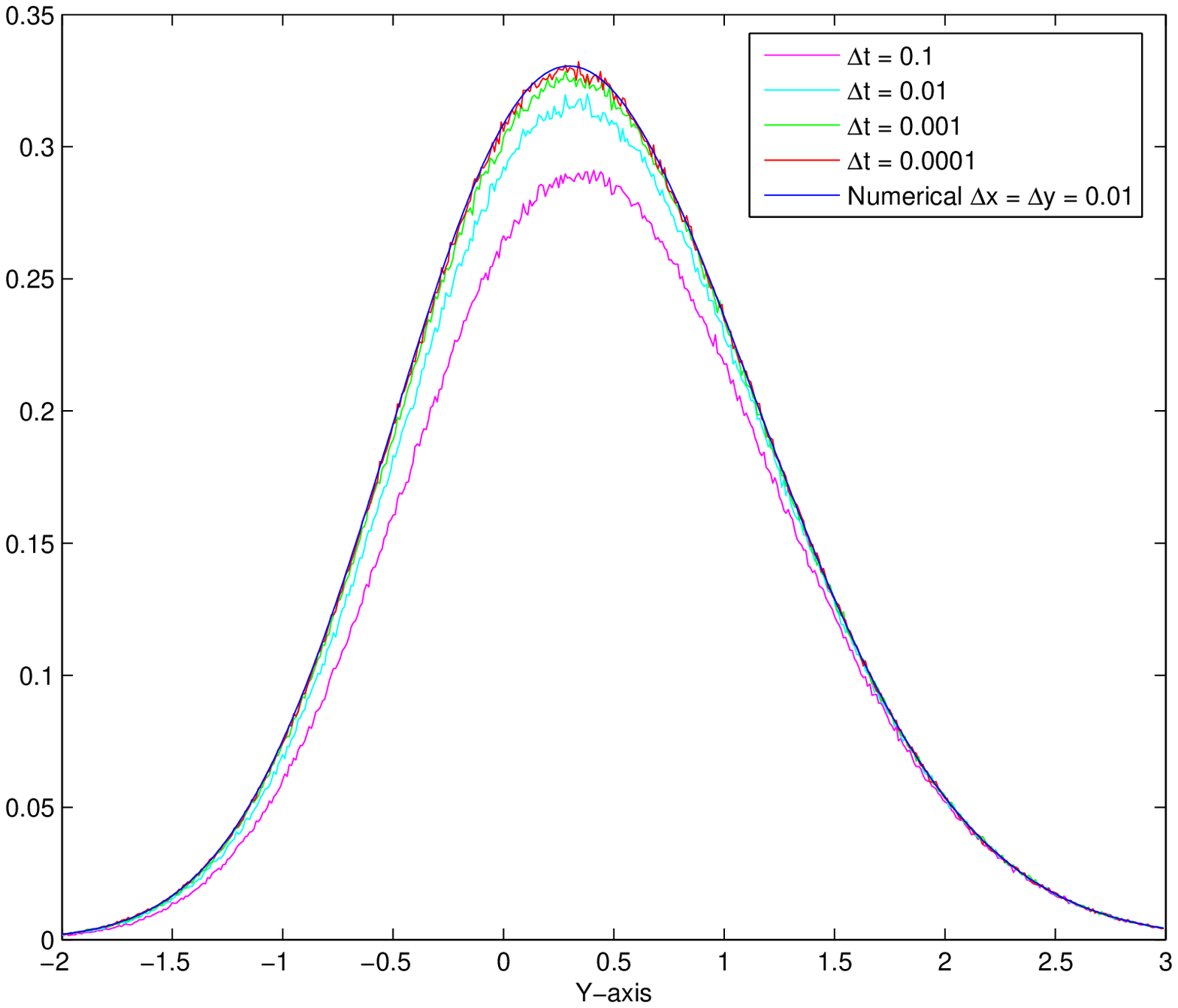}} \caption {The
marginal density of $y(T)$ with no drift and correct oblique
reflection (the first experiment). The numerical solution of the FPE
(blue) with grid size $\Delta x = 0.01$ and estimates from the
simulation of $n = 10^7$ trajectories with time steps $\Delta t =
10^{-1}, 10^{-2}, 10^{-3}, 10^{-4}$. }
\label{fig:implicit_correct_y}
\end{center}
\end{figure}

In the first experiment the drift vanishes, $\mb{a}=0$. The normal
$\mb{n}=(1,0)$ and the co-normal $\mb{\sigma}\mb{n}=(0.25, 0.4)$
point in different directions. The simulated trajectories are
reflected in the co-normal direction according to the prescription
(\ref{co-normal}). The simulated and the numerical solution of the
FPE give the marginal densities shown in Figures
\ref{fig:implicit_correct_x} and \ref{fig:implicit_correct_y}.
Figure \ref{fig:implicit_correct_x} shows the marginal density of
$x(T)$,
\begin{equation*}
p(x,T\,|\,x_0,y_0) = \int_{-\infty}^\infty p(x,y,T\,|\, x_0,y_0) \,dy,
\end{equation*}
while Figure \ref{fig:implicit_correct_y} shows the marginal density
of $y(T)$,
\begin{equation*}
p(y,T\,|\,x_0,y_0) = \int_{0}^\infty p(x,y,T\,|\, x_0,y_0) \,dx.
\end{equation*}
Table~\ref{t:surv_n_1} gives the computed survival probability and
indicates the convergence rate.

\begin{figure} [htbp]
\begin{center}
{\epsfxsize=400pt\epsffile{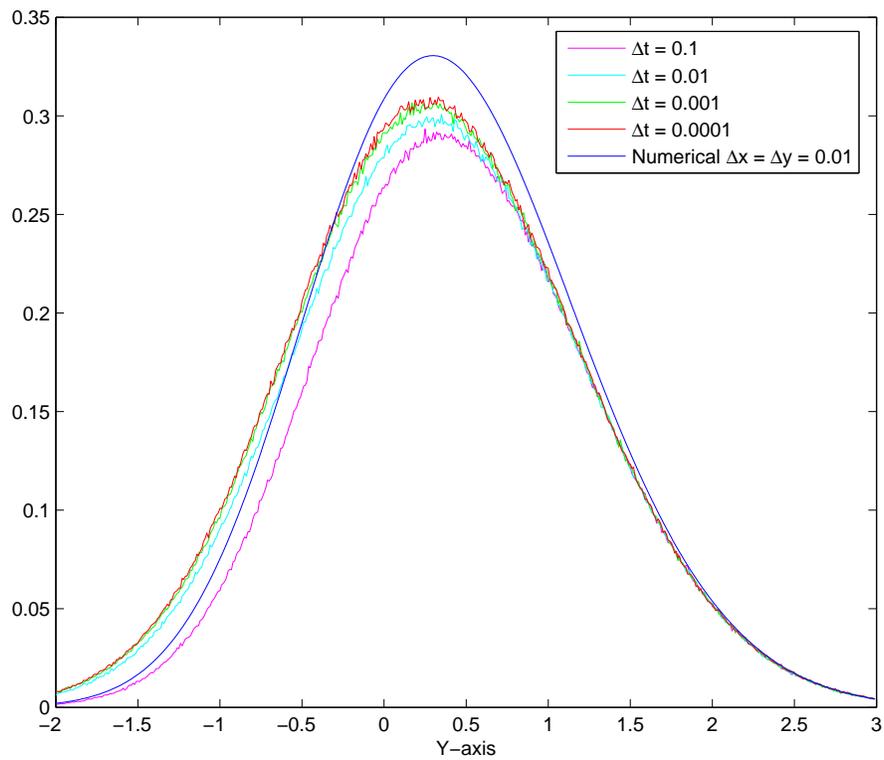}} \caption {The
marginal density of $y(T)$ with no drift and with normal reflection
(the second experiment). The numerical solution of the FPE (blue)
with grid size $\Delta x = 0.01$ and estimates from the simulation
of $n = 10^7$ trajectories with time steps $\Delta t = 10^{-1},
10^{-2}, 10^{-3}, 10^{-4}$. } \label{fig:implicit_wrong_y}
\end{center}
\end{figure}

We illustrate the importance of using the correct reflection law in
the second experiment, in which the simulated trajectories are
reflected in the normal direction $\mb{n}=(1,0)$. Clearly, the
marginal density of $x(T)$ coincides with that of the first
experiment, because both oblique and normal reflections have the
same $x$-coordinate (see (\ref{Euler-n2})). However, the plot of the
marginal density of $y(T)$ differs significantly from that in the
previous experiment. It is apparent from the comparison to the
numerical solution of the FPE that the simulation does not recover
the Robin boundary condition in the limit $\Delta t\to0$ (see
Figure~\ref{fig:implicit_wrong_y}). Note that the peak of the
density is at $y>0$, though the reflection is normal. This is due to
the anisotropy of the diffusion tensor, which causes the probability
flux density vector to have a positive $y$ component.

\begin{figure} [htbp]
\begin{center}
{\epsfxsize=400pt\epsffile{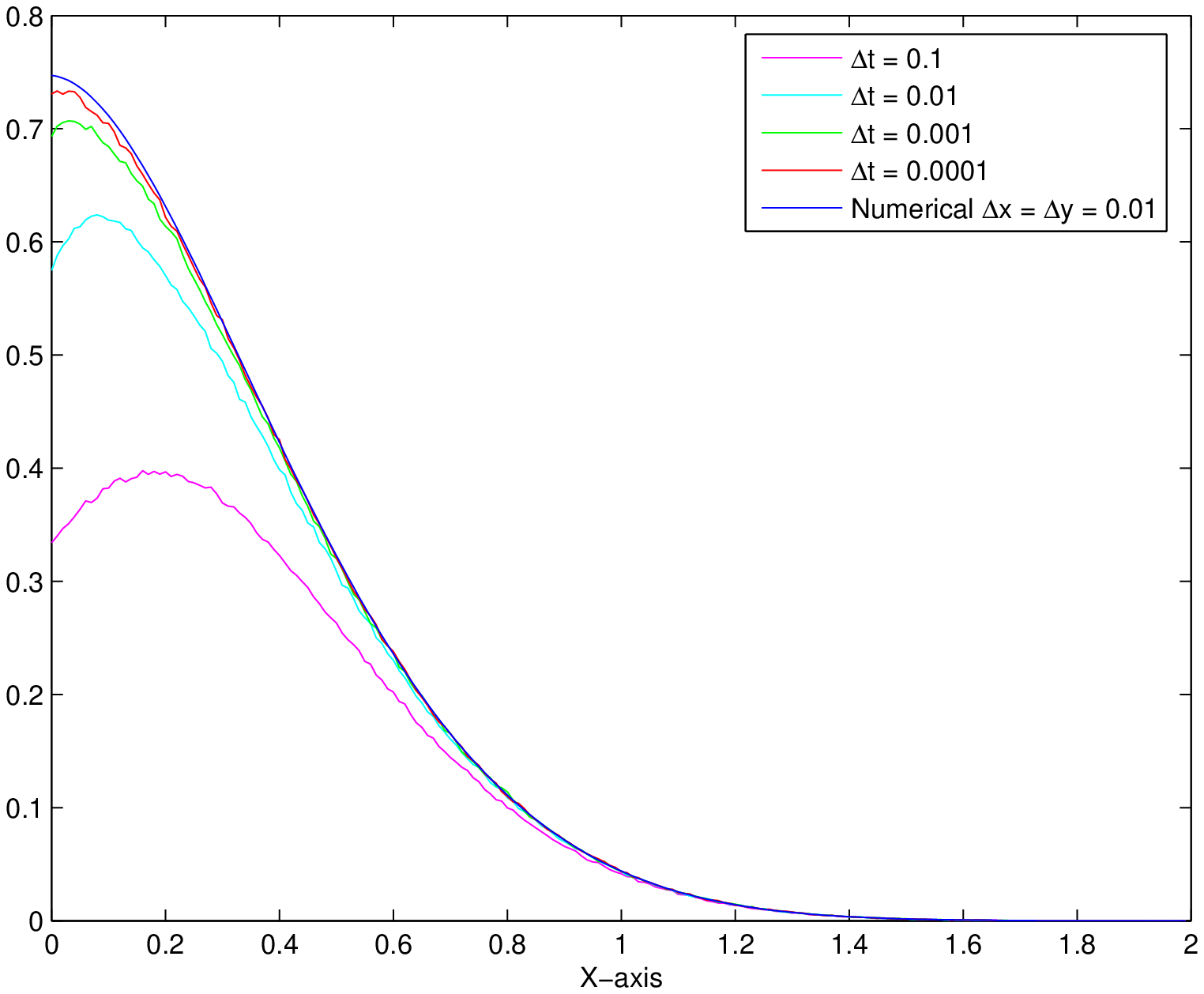}} \caption {The
marginal density of $x(T)$ with drift $\A = (-1, 0)$ and correct
oblique reflection (the third experiment). The numerical solution of
the FPE (blue) with grid size $\Delta x = 0.01$ and estimates from
the simulation of $n = 10^7$ trajectories with time steps $\Delta t
= 10^{-1}, 10^{-2}, 10^{-3}, 10^{-4}$.} \label{fig:implicit_drift_x}
\end{center}
\end{figure}

\begin{figure} [htbp]
\begin{center}
{\epsfxsize=400pt\epsffile{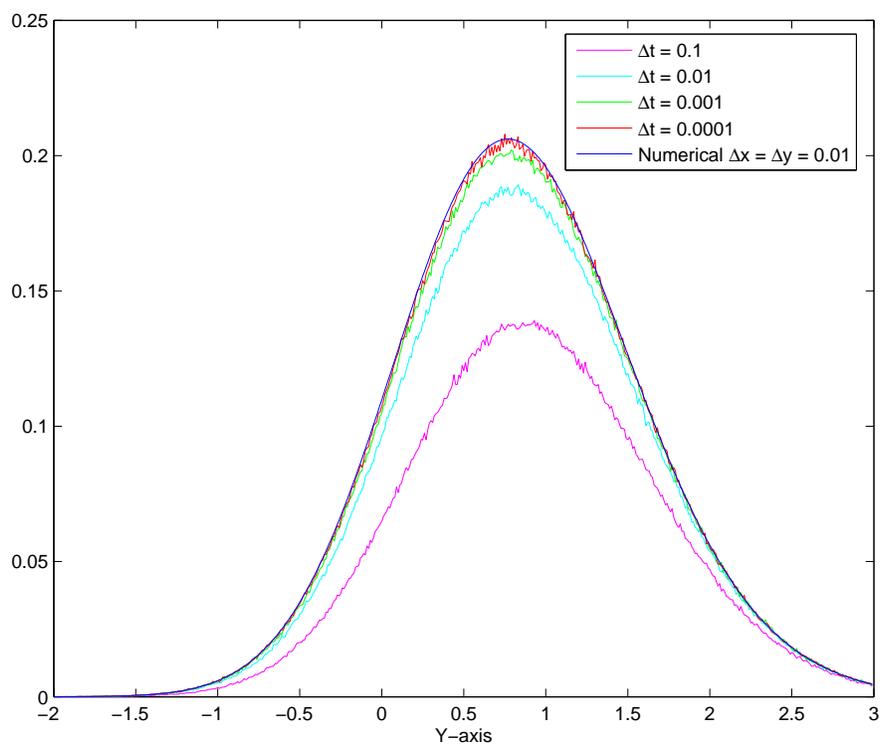}} \caption {The
third experiment ($\A = [-1, 0]^T$, correct oblique reflection):
y-marginal densities. The numerical solution (blue) is compared to
four simulated ones (with time steps $\Delta t = 10^{-1}, 10^{-2},
10^{-3}, 10^{-4}$). $n = 10^7$. Resolution: $\Delta x = 0.01$. }
\label{fig:implicit_drift_y}
\end{center}
\end{figure}

In the third experiment the drift is the constant vector $\A  =
\left(-1,0\right)$ and the diffusion tensor is as in the first
experiment. The density is shifted toward the boundary (see
Figure~\ref{fig:implicit_drift_x} and
Figure~\ref{fig:implicit_drift_y}). The results are summarized in
Table~\ref{t:surv_n_3}.

\section{\label{sec:discussion}Summary and Discussion}

We have defined a diffusion process with partially reflecting
boundary as a limit of Markovian jump processes generated by the
Euler scheme for the dynamics in a half space with partial
absorption  of exiting trajectories and partial oblique reflection
in the boundary. We derived an expression for the radiation constant
in the Robin boundary condition for the one-dimensional
Fokker-Planck equation for the case of diffusion with variable drift
and diffusion coefficients, as a function of the absorption
probability. We found that the Euler scheme for a diffusion in a
half space with variable drift and constant anisotropic diffusion
has to be reflected in a particular oblique direction in order to
recover the Robin boundary condition. Also for this case we found
the radiation "constant" as a function of the local absorption
probability on the boundary. We found a boundary layer of width
$O(\sqrt{\Delta t})$ in the pdf of the Euler scheme and solved the
boundary layer equation, which is of Wiener-Hopf type.

The boundary layer of $p_{\Delta t}(y,t)$ makes the calculation of
the boundary flux non-trivial. The net boundary flux of the
simulation profile $p_{\Delta t}(y,t)$ is
 \beq
 -J_{\Delta t}(0,t)=\lim_{\Delta t\to0}\frac{1}{\Delta t}
 \frac{P\sqrt{\Delta t}}{\sqrt{4\pi\sigma\Delta
 t}}\int_{-\infty}^0dy\int_0^\infty
 p_{\Delta t}(x,t)\exp\left\{-\frac{(x-y)^2}{4\pi\sigma\Delta
 t}\right\}\,dx,\label{Jint}
 \eeq
which is the probability of the trajectories that propagate per unit
time out of the domain, discounted by the probability of
trajectories returned into the domain by the partially reflecting
Euler scheme. Changing  the order of integration and then changing
the variable of integration into $z=x/2\sqrt{\sigma\Delta t}$ gives
 \beq
 -J_{\Delta t}(0,t)=P\sqrt{\sigma}\int_0^\infty\mbox{erfc}(z)p_{\Delta
 t}(2z\sqrt{\sigma\Delta t},t)\,dz=\frac{P\sqrt{\sigma}}{\sqrt{\pi}}
 p^{(0)}_{BL}(0,t)+O(\sqrt{\Delta t}).\label{referee}
 \eeq
This straightforward calculation of the flux gives the correct
radiation constant, provided that
 \beq
 p^{(0)}_{BL}(0,t)=p_{OUT}^{(0)}(0,t).\label{provision}
 \eeq
The latter, however, depends on the mode of reflecting a trajectory
from $x'$ outside to $x''$ inside the domain. We have shown that for
$x''=-x'$  the provision holds, however, for other schemes, e.g.,
$x''=-\alpha x'$ ($\alpha\neq1$), the provision (\ref{provision})
fails in general, though (\ref{referee}) still holds. On the other
hand, the differential form of the flux, (\ref{J}), has to be
obtained from (\ref{Jint}) in the limit $\Delta t\to0$, which is not
the case for $p_{\Delta t}(y,t)$, though it is for $p_{OUT}(y,t)$.
This shows up in spades in the multi-dimensional case, because
although (\ref{provision}) holds for any direction of reflection,
yet the differential form of the flux is obtained in the limit only
if the correct direction of oblique reflection is chosen.

The generalization of the multi-dimensional case to domains with
curved boundaries and to a variable diffusion tensor
$\mb{\sigma}(\x,t)$ is not straightforward and will be done
separately. Note that if the diffusion tensor is constant, but
un-isotropic, a local orthogonal mapping of the boundary to a plane
converts the diffusion tensor from constant to variable, as can be
seen from It\^o's formula. However, as mentioned in Section
\ref{sec:intro}, in the most common case of constant isotropic
diffusion, our result extends to domains with curved boundaries,
because the mapping leaves the Laplacian unchanged, though the drift
changes according to It\^o's formula. In this case the vector $\U$
coincides with the normal $\n$.

\newpage

\begin{table}[h]
\begin{center}
\begin{tabular}{c|c|c}
$\Delta t$ & $n_{sur}$ & $p_{sur}-n_{sur}/n$\\
\hline $10^{-1}$ & 7253450 & 0.0456\\
$10^{-2}$ & 7577156 & 0.0132 \\
$10^{-3}$ & 7670969 & 0.0039 \\
$10^{-4}$ & 7698523 & 0.0011
\end{tabular}
\end{center}
\caption{Survival probability: the difference between the analytic
value of the survival probability $p_{sur}=0.77095\ldots$ and its
numerical estimation $n_{sur}/n$ decreases by roughly $\sqrt{10}$
whenever $\Delta t$ is decreased by an order of magnitude.
(Parameters: $\sigma=\kappa=x_0=t=1$, $a=0$, $n=10^7$)}
\label{t:surv}
\end{table}
\begin{table}[htbp]
\begin{center}
\begin{tabular}{c|c|c}
$\Delta t$ & $n_{sur}$ & $p_{sur}-n_{sur}/n$ \\
\hline
$10^{-1}$ & 5986662 & 0.0814708 \\
$10^{-2}$ & 6449991 & 0.0351379 \\
$10^{-3}$ & 6707318 & 0.0094052 \\
$10^{-4}$ & 6775672 & 0.0025698 \\
\end{tabular}
\end{center}
\caption{Survival probability for $\A=\mb{0}$. The third column
lists the error between the numerical value of the survival
probability $p_{sur}=0.6799545$ from the solution of the FPE and its
estimate $n_{sur}/n$ from the simulation. The error decreases by
about $\sqrt{10}$ whenever $\Delta t$ is decreased by an order of
magnitude, indicating the convergence rate $\sqrt{\Delta t}$ of the
simulation.} \label{t:surv_n_1}
\end{table}
\begin{table}[htbp]
\begin{center}
\begin{tabular}{c|c|c}
$\Delta t$ & $n_{sur}$ & $ p_{sur}-n_{sur}/n$ \\
\hline
$10^{-1}$ & 2541947 & 0.1180946                 \\
$10^{-2}$ & 3399528 & 0.0323365 \\
$10^{-3}$ & 3632622 & 0.0090271 \\
$10^{-4}$ & 3693905 & 0.0028988 \\
\end{tabular}
\end{center}
\caption{Survival probability for $a=(-1,0)$. The third column lists
the error between the numerical value of the survival probability
$p_{sur}=0.3722893$ from the solution of the FPE and its estimate
$n_{sur}/n$ from the simulation.} \label{t:surv_n_3}
\end{table}
\newpage

\end{document}